\pdfoutput=1
\documentclass[aps, reprint, superscriptaddress, letterpaper, floatfix, prb]{revtex4-1}
\usepackage{graphicx}
\usepackage{upgreek}
\usepackage{color}
\usepackage{subfigure}
\usepackage{tabularx}
\usepackage{floatrow}
\usepackage{dcolumn}
\usepackage{bm}
\usepackage{amsmath}
\usepackage{scrextend}
\usepackage[normalem]{ulem}
\usepackage{hyperref}

\begin{document}

\title{Bloch equations in Terahertz magnetic-resonance ellipsometry}

\author{Viktor Rindert}
\email[Electronic mail: ]{viktor.rindert@ftf.lth.se}
\affiliation{NanoLund and Solid State Physics, Lund University, S-22100 Lund, Sweden}
\affiliation{Terahertz Materials Analysis Center, THeMAC, Lund University, S-22100 Lund, Sweden}
\affiliation{Center for III-Nitride Technology, C3NiT - Janz\'en, Lund University, S-22100 Lund, Sweden}

\author{Steffen Richter}
\affiliation{NanoLund and Solid State Physics, Lund University, S-22100 Lund, Sweden}
\affiliation{Terahertz Materials Analysis Center, THeMAC, Lund University, S-22100 Lund, Sweden}

\author{Philipp Kühne}
\affiliation{Department of Physics, Chemistry, and Biology (IFM), Link{\"o}ping University, SE 58183, Link{\"o}ping, Sweden}
\affiliation{Terahertz Materials Analysis Center, THeMAC, Link{\"o}ping University, SE 58183, Link{\"o}ping, Sweden}
\affiliation{Center for III-Nitride Technology, C3NiT - Janz\'en, Link{\"o}ping University, SE 58183, Link{\"o}ping, Sweden}

\author{Alexander Ruder}
\affiliation{Department of Electrical and Computer Engineering, University of Nebraska-Lincoln, Lincoln, NE 68588, USA}

\author{Vanya~Darakchieva}
\affiliation{NanoLund and Solid State Physics, Lund University, S-22100 Lund, Sweden}
\affiliation{Terahertz Materials Analysis Center, THeMAC, Lund University, S-22100 Lund, Sweden}
\affiliation{Center for III-Nitride Technology, C3NiT - Janz\'en, Lund University, S-22100 Lund, Sweden}
\affiliation{Department of Physics, Chemistry, and Biology (IFM), Link{\"o}ping University, SE 58183, Link{\"o}ping, Sweden}

\author{Mathias Schubert}
\affiliation{NanoLund and Solid State Physics, Lund University, S-22100 Lund, Sweden}
\affiliation{Terahertz Materials Analysis Center, THeMAC, Lund University, S-22100 Lund, Sweden}
\affiliation{Department of Electrical and Computer Engineering, University of Nebraska-Lincoln, Lincoln, NE 68588, USA}

\date{\today}

\begin{abstract}
{A generalized approach derived from Bloch’s equation of motion of nuclear magnetic moments is presented to model the frequency, magnetic field, {spin} density, and temperature dependencies in the electromagnetic permeability tensor for materials with magnetic resonances. The resulting tensor model predicts characteristic polarization signatures which can be observed, for example, in fully polarization-resolved Mueller matrix element spectra measured across magnetic resonances as a function of frequency, magnetic field, {magnetic} moment density, and temperature. When augmented with thermodynamic considerations and suitable Hamiltonian description of the magnetic eigenvalue spectrum, important parameters such as zero-frequency magnetization, spectral amplitude distribution, relaxation time constants, and geometrical orientation parameters of the magnetic moment density can be obtained from comparing the generalized model approach to experimental data. We demonstrate our approach by comparing model calculations with full Mueller matrix element spectra measured at oblique angle of incidence in the terahertz spectral range, across electron spin resonance quintuplet transitions observed in wurtzite-structure GaN doped with iron. Measurements were performed by ellipsometry, using a superconducting cryostat magnet at magnetic fields of $\pm$7.23~T and at temperatures of 20~K and 30~K. We detail the occurrence of linear and circular birefringence and dichroism associated with each of the zero-field split spin transitions in the $S=5/2$ defect system. We derive the spectral dependence of the magnetic susceptibility function and obtain the temperature and magnetic field dependence of the spin Hamiltonian. Our model correctly predicts the complexity of the polarization signatures observed in the 15 independent elements of the normalized Mueller matrix for both positive and negative magnetic fields, and also permits detailing the orientation of the magnetic moments at the Fe defect sites. Our model will be useful for future analysis of frequency and magnetic field-dependent magnetic resonance measurements.}

\end{abstract}

\maketitle

\section{Introduction}

Magnetic Resonance Spectroscopy stands as a cornerstone in the realm of molecular and material{s} research. Originating from the seminal observations by Zeeman on the impact of magnetic fields on spectral lines, Magnetic Resonance Spectroscopy has evolved into a comprehensive suite of techniques, including Nuclear Magnetic Resonance (NMR), Electron Paramagnetic Resonance (EPR), and Ferromagnetic Resonance (FMR). These methodologies have {provided} unparalleled insights into molecular structures, and dynamics, significantly advancing our understanding across chemistry, materials science, medicine, and semiconductor physics. EPR is widely used to study quantum transitions of molecular magnetic moments.\cite{Poole1983} Conventional EPR instruments detect magnetic resonance at a few fixed frequencies, typically at a few tens of GHz.\cite{doi:10.1021/ja003707u} The Zeeman splitting energy, $\Delta E$, of a free electron at 10~GHz, for example, corresponds to a magnetic field of $B=\frac{\Delta E}{g\mu_{B}}=357$~mT, and vacuum wavelength of $\lambda \approx3$~cm, where $\mu_B=\frac{e\hbar}{2m_e}$ is the Bohr magneton, $e$ is the unit electric charge, $\hbar$ is the reduced Planck constant, and $m_e$ is the free electron mass. Pioneered by Felix Bloch in his {seminal} 1946 paper,\cite{PhysRev.70.460} electromagnetic absorbance loss measurements of magnetic transitions are {typically} performed in resonator cavities at fixed frequencies under near-field conditions and slowly {scanning} external magnetic field permitting for relatively small sample size and highly sensitive detection. Historically, thus conventional EPR practice is positioning the sample within a resonant microwave cavity. While effective for certain applications, this setup inherently imposes a restriction on the system, confining it to a finite set of discrete frequencies. This limitation has traditionally hindered the ability to fully explore frequency-dependent {magnetic suscetibility} characteristics.\cite{Savitsky2009} Tuning frequency or modifying polarization in such resonator geometries is difficult, however, investigating magnetic resonances at larger magnetic fields is generally desirable since energy and magnetic field resolutions improve. At 100~GHz, $\lambda \approx 3$~mm and a field of $B=3.57$~T are needed for the single free electron Zeeman splitting to match. Such fields can be conveniently achieved with superconducting magnets. Studies of magnetic properties at higher magnetic fields are of interest for large and complex magnetic systems,\cite{C7CP07443C,LAGUTA2018138} for systems with large spin quantum numbers and strong spin-spin coupling,\cite{doi:10.1021/ja00139a021,HASSAN2000300} {and} for more comprehensive investigations of defect properties in wide-bandgap semiconductors, e.g., in SiC\cite{ PhysRevB.103.245203, doi:10.1063/1.4866331, SonAMR2010} {and GaN,} and in ultrawide-band-gap semiconductors, e.g., {AlN}\cite{ ZvanutJEM2019, Sunay_2019}, or gallium oxide.\cite{doi:10.1063/1.4990454,doi:10.1063/1.5053158,doi:10.1063/1.4972040,doi:10.1063/5.0002763, doi:10.1063/1.5133051,doi:10.1063/1.5127651,doi:10.1063/1.5081825,doi:10.1063/5.0012579} 

In recent years, several groups have developed new methods of performing EPR at higher fields with the possibility of tuning the frequency used.\cite{2017tds, Neugebauer, 2005Krzystek, 2022Bray, 2021fepr} Transitioning the resonance conditions into the THz range, far-field optical techniques such as spectroscopic ellipsometry become available. In addition to frequency variation with few-kHz resolution capability, {frequency multiplication of} highly polarized solid-state synthesizer sources have become available recently as well. Implementing suitable polarization modulation, THz spectroscopic ellipsometry was developed lately, permitting measurement of Mueller matrix {spectra}.\cite{doi:10.1063/1.4889920,2018LuEllipsometer} Thereby, the anisotropic THz optical properties of samples with complex layer structures and low-symmetry crystal structures can be assessed and quantified. THz ellipsometry was also demonstrated on samples immersed in strong magnetic fields within superconducting split-coil Helmholtz-type magnets for measurement of the optical Hall effect. \cite{doi:10.1063/1.4889920,2018LuEllipsometer,KnightRSI2020} Such measurements can be performed in reflection and transmission geometries and have demonstrated sensitivity to free charge carrier properties such as effective mass, mobility, density, and {charge} (electron, hole). Employing the advantages of free space propagation such as freely tunable frequency, free choice of propagation direction, and, most importantly, {variation of} all possible polarization conditions, THz EPR ellipsometry permits investigations of magnetic transitions, including their energy-field diagrams as well as their polarization properties.\cite{SiCpaper,2022MORoadmap} THz EPR thereby {dispenses} with the need for resonance cavities and permits the study of semiconductor heterostructures and thin films. Frequency-field maps allow investigators to quickly identify the origins of observed transitions because different mechanisms, such as hyperfine structure splitting, Zeeman splitting, or zero-field splitting (ZFS), reveal different frequency-field diagrams. Also, level populations change with field and, therefore, with frequency and reveal dynamics in the complex-valued spin susceptibility, which can be measured using ellipsometry.\cite{Poole1983,WeilBolton2007}

The previously developed THz Optical Hall effect instrumentation is readily available for THz EPR ellipsometry {measurements}. Field and frequency dependent linearly and circularly polarized Mueller matrix element spectra were measured of the magnetic spin transitions associated with cubic and hexagonal lattice site nitrogen defects in 4H-SiC.\cite{SiCpaper} {In our recent work, we have} reported {field-scanning} THz-EPR ellipsometry {measurements at selected frequencies} of transitions associated with iron defects in $\beta$-Ga$_2$O$_3$. {\cite{RichterPRBFebGO2024}} The high-spin $S = 5/2$ state of Fe$^{3+}$ causes five transitions which were all detected simultaneously, in contrast to conventional low-field EPR. Analysis of the ellipsometry data, {based on direct lineshape matching to
the Mueller-matrix elements,} resulted in the full set of fourth-order monoclinic zero-field splitting parameters for the octahedrally and tetrahedrally coordinated defects. A subsequent Hamiltonian analysis revealed that simplified second-order orthorhombic
approximations are insufficient to model the high spin system at high magnetic fields owing to the monoclinic symmetry of the host crystal.\cite{RichterPRBFebGO2024}

Ellipsometry is {a} method for accurately characterizing the linear optical properties of materials, particularly of thin films.\cite{Fujiwara} Data analysis requires model calculations which consider plane wave propagation throughout the sample including Fabry{-}Perot interference, and requires appropriate assumptions about the complex-valued dielectric ($\varepsilon$) and magnetic ($\mu$) {material} functions within  Maxwell's equations, $\mathbf{D}=\varepsilon_0\varepsilon\mathbf{E}$, and $\mathbf{B}$=$\mu_0\mu\mathbf{H}$, where $\mathbf{D}$ and $\mathbf{E}$, and $\mathbf{B}$ and $\mathbf{H}$ are dielectric displacement and electric field phasors, and magnetic induction and magnetic field phasors, respectively. Both response functions are second-rank tensors and are complex-valued. The structure of the tensors reveals the underlying physical polarization processes that lead to the optical response of a given system. It is commonly accepted for optical frequencies that $\mu$ is unity and isotropic, independent of frequency and wave vector. However, at THz frequencies across magnetic transitions, the magnetic susceptibility can differ significantly from unity. This was demonstrated in our previous work reporting the field- and frequency-scanning THz-EPR ellipsometry measurement of the Nitrogen defect in SiC.\cite{SiCpaper} In order to perform model analysis to obtain accurate quantitative physical parameters in ellipsometry, a suitable model must be selected. For dielectric processes, such models are referred to as model dielectric function (MDF) approaches which are then used to identify, for example, band-to-band transition energies, or phonon modes and their eigendielectric polarization directions.\cite{PhysRevB.93.125209,PhysRevB.95.165202,Fujiwara} These {MDF} models are derived and founded on physical models and fundamental principles of classical and quantum mechanics including energy, charge, and momentum conservation. However,~\textit{ad-hoc} model functions are often used, which lack rigorous physical derivations. For analysis of our previous THz-EPR ellipsometry data,\cite{SiCpaper} such an~\textit{ad-hoc} model was implemented. This model sufficed to reproduce the main features of the ellipsometry spectra. However, other characteristic measured properties did not correspond well with the~\textit{ad-hoc} model. Previous model descriptions that correctly render the complex-valued frequency and external magnetic field dependencies for the tensor appearance of the magnetic susceptibility for spin or general magnetic transitions are not known from ellipsometry investigations. Hence, such model descriptions must be searched for and tested. In this paper, we present a derivation of a magnetic permeability tensor that describes the full magneto-optic response of a material and sufficiently explains THz EPR ellipsometry data for spin transitions. We derived this model from the Bloch equations\cite{PhysRev.70.460} of the magnetic induction under the influence of an external static and high-frequency magnetic field. We believe this model will also become useful for analysis of the frequency and field dependent responses of general magnetic transitions and excitations.

The Bloch equations\cite{PhysRev.70.460} describe the time evolution of a macroscopic ensemble of magnetic moments by phenomenologically introducing relaxation times. The 'Bloch susceptibility' is the form of the magnetic susceptibility derived from the Bloch equations. As the signal strength is proportional to the magnetization, the imaginary part of the Bloch susceptibility is often used to describe absorption spectra within the fields of EPR and nuclear magnetic resonance (NMR).\cite{Gutowsky1954, 1958McConnell, 2000Lineshapes, 1975PulsedEPR, 1969Lineshape,1994lineshape} By deconvoluting the lineshape, the relaxation times of the studied magnetic moments can be {obtained}. These relaxation times are intricately linked to the local interactions of paramagnetic centers with their surrounding environment{, and yield} information regarding the studied system's electronic structure and molecular motions. \cite{Eaton2000} The absorption spectra {obtained} by conventional EPR are in arbitrary units. {Analysis of information related to absolute} amplitude requires great care. This can be done by using a reference sample with a known spin volume concentration or by studying polarization {effects} {rather than intensity} as performed in this study. Once obtained, the amplitude of {an EPR signature} can be related to the longitudinal relaxation time and the spin volume concentration.

{Our work here} aims to rigorously evaluate the efficacy of employing the Bloch equation formalism for interpreting magnetic resonance data, specifically obtained through Mueller matrix ellipsometry.\cite{MuellerMIT1943,Fujiwara,HilfikerHongSchoeche+2022+59+91,Schubert:16} The original Bloch equations, while proficient in providing a fundamental and intuitive framework for understanding the rate processes inherent in magnetic resonance phenomena, frequently fall short in precisely replicating experimental outcomes. This discrepancy arises primarily due to the oversimplified nature of the phenomenological relaxation times, which inadequately capture the complex dynamics of magnetic moments, particularly in scenarios where there is significant interaction and interconversion among them. Here, we demonstrate the validity of the Bloch model by robust agreement between measured and best-match model calculated Mueller matrix spectra for defect-induced electron spin resonance transitions in Fe-doped GaN as an example. The zero-field splitting parameters of the $S=5/2$ high-spin defect Fe$^{3+}$ in GaN have been investigated previously with fixed frequency low-field EPR.\cite{Kashiwagi_2007} The frequency-dependent measurements at fixed magnetic fields revealed in this work demonstrate the new capabilities of THz EPR ellipsometry to directly infer spin parameters from {individual spectra}. In addition, the spin Hamiltonian parameters can be determined from analysis of the eigenvalue spectrum at fixed magnetic fields.

In Sec.~\ref{Sec:BlochSusceptibility}, we present the derivation of the frequency-dependent {complex-valued} Bloch susceptibility {tensor}, which is then expanded to high spin systems in Sec.~\ref{sec:BlochHighSpin}. Based on the results in Sec.~\ref{Sec:BlochSusceptibility} we demonstrate the structure of the anisotropic magnetic response function tensor and the resulting anisotropic properties. We thereby introduce the Bloch eigenmagnetic polarizability functions and eigenvectors in Sec.~\ref{sec:BlochEigenpolModel}. These functions establish the base of the Model Magnetic Function (MMF) approach used in Sec.~\ref{sec:MMFparameteranalysis} for analysis of THz-EPR ellipsometry measurements performed on Fe-doped GaN. In Sect.~\ref{Method:EPR} we present our THz-EPR ellipsometry method used here. Data reduction and parameter determination are discussed in Sec.~\ref{sec:MMFparameteranalysis2}. Section~\ref{sec:Fe3+EPR} details the results of the analysis for low-density Fe-doped GaN. Lineshape analysis results, including amplitude, broadening, and frequency parameters of the two equivalent GaN lattice sites' zero-field-split spin transition quintuplets, are shown, discussed, and compared with available literature data from low-field, fixed-frequency EPR. We then demonstrate the quantitative determination of the spin density using the results of the MMF analysis in combination with a Brillouin magnetization summation approach. The experiment on the spin $S=5/2$ system in GaN is selected for the purpose of demonstration. The Hamiltonian description of the zero-field split system is well known and, thereby, is suitable for the purpose here. For this reason, the spin measurements are performed in a specific orientation of the crystal such that Fe$^{3+}$ in two equivalent sites produce identical spin eigenvalues, and thereby simplify the analysis {for improved transparency of our MMF approach implemented here}.

\section{Theory}

\subsection{Bloch susceptibility}\label{Sec:BlochSusceptibility}
To predict the frequency dependence of the complex-valued permeability tensor for a set of magnetic resonances, a model based on the Bloch equations is derived. The set of Bloch equations describes how a magnetization vector $\textbf{M}=(M_x, M_y, M_z)$ responds to an external perturbation when subjected to a static or slow-changing magnetic induction field $B_0$. {The magnetic moment associated with the magnetization vector is thought of a continuum density, and may be interpreted as caused by a homogeneous distribution of infinitely small magnetic objects with volume (spin) density $n_e$.} Without loss of generality, with $B_0$ oriented along direction $z$, the Bloch equations including the phenomenological relaxation times are represented as follows\cite{Slichter1990, boltonchapter10}
\begin{align}
    \begin{split}
       & \frac{\partial M_x}{\partial t} = -\omega_0 M_y - \frac{M_x}{T_2},\\
       & \frac{\partial M_y}{\partial t} = \omega_0 M_x - \frac{M_y}{T_2},\\
       & \frac{\partial M_z}{\partial t} = - \frac{M_z - M_0}{T_1},
    \end{split}
    \label{Bloch1}
\end{align}
where $\omega_0=\gamma_e B_0$ is the classical Larmor frequency, $\gamma_e$ is the gyromagnetic ratio, $M_0$ is the {dc} magnetization {along the z-axis and is assumed to be linearly proportional to $B_0$}, $T_1$ is the longitudinal relaxation time, and $T_2$ is the transverse relaxation time. To assess the response of vector $\mathbf{M}$ to an oscillating magnetic induction field, i.e., a time-harmonic external perturbation due to an electromagnetic plane wave traversing a magnetized medium, it is advantageous to employ a rotating frame of reference with an identical angular frequency $\omega$ as that of the oscillating field. Moreover, assuming that the oscillating field is comprised of left-circularly polarized (LCP) light whose oscillating amplitude $H_1=\frac{1}{\mu_0}B_1$ ({CGS units are used throughout this derivation, resulting in $H_1=B_1$)} then remains fixed along the $x$-axis within the rotating frame, where $\mu_0$ is the vacuum permeability and $H_1$ ($B_1$) is the auxiliary magnetic (induction) field, the Bloch equations transform into 
\begin{align}
    \begin{split}
        & \frac{\partial \Tilde{M_x}}{\partial t} = -(\omega_0 - \omega) \Tilde{M_y} - \frac{\Tilde{M_x}}{T_2},\\
       & \frac{\partial \Tilde{M_y}}{\partial t} = (\omega_0 - \omega) \Tilde{M_x} - \frac{\Tilde{M_y}}{T_2} - \gamma_e B_1 M_z  ,\\
       & \frac{\partial M_z}{\partial t} = - \frac{M_z - M_0}{T_1} + \gamma_e B_1 \Tilde{M_y} ,
    \end{split}
    \label{Bloch2}
\end{align}
where ``$^\sim$'' annotates components of $\mathbf{M}$ in the rotating frame around the $z$-axis. Solving \eqref{Bloch2} for a steady-state expression while assuming very large longitudinal relaxation time $T_1$ yields\cite{Slichter1990}
\begin{align}
    \begin{split}
    & \Tilde{M_x} = \chi_0  T_2 \omega_0  \frac{  (\omega_0 - \omega)T_2 }{1 + T_2^2(\omega_0 -\omega)^2}B_1,\\
    & \Tilde{M_y} =  \chi_0  T_2 \omega_0 \frac{1}{1 + T_2^2(\omega_0 -\omega)^2}B_1,\\
    & M_z = M_0,
    \end{split}
    \label{chi_def}
\end{align}
where we have assumed that $B_1<<B_0$, which is typically the case, and we introduce the {dc} susceptibility $\chi_0= M_0/B_0$. We can thus express frequency-dependent susceptibility functions $\chi'$ and $\chi''$ in the rotating frame of reference
\begin{align}
    \begin{split}
        &\chi'(\omega) = \frac{\Tilde{M_x}}{B_1} = \chi_0  T_2 \omega_0  \frac{  (\omega_0 - \omega)T_2 }{1 + T_2^2(\omega_0 -\omega)^2},\\
        &\chi''(\omega) =\frac{\Tilde{M_y}}{B_1} = \chi_0  T_2 \omega_0 \frac{1}{1 + T_2^2(\omega_0 -\omega)^2}.
    \end{split}
\end{align}
We consider the magnetization response to a left-handed circularly polarized (LCP, +) electromagnetic plane wave and seek the complex-valued frequency-domain Bloch susceptibility $\chi_{B,+}\equiv \chi_1 + i\chi_2$ in the static reference frame. The amplitude of the oscillating induction component of the electromagnetic field in the $x$-direction can be expressed as $B_x=B_1e^{-i\omega t}$ {when} the plane wave is propagating along the positive direction of the \textit{z}-axis. Consequently,
\begin{align}
    \begin{split}
        &M_x = \Re{(\chi_{B,+}B_1e^{-i\omega t})} = \\& B_1(\chi_1 \cos(\omega t) + \chi_2 \sin (\omega t)),
    \end{split}
    \label{Mx}
\end{align}
where $\Re$ denotes the real part. We then express the magnetization through the components in the rotating frame
\begin{align}
    \begin{split}
        &M_x = \Tilde{M_x} \cos{\omega t} + \Tilde{M_y} \sin (\omega t) =\\
        &B_1(\chi ' \cos (\omega t) + \chi'' \sin (\omega t)),
    \end{split}
\end{align}
and by comparison with Eq.~\eqref{Mx}
\begin{equation}
    \chi_{B,+} = \chi'(\omega) + i \chi''(\omega),
\end{equation}
or equivalently
\begin{equation}
    \chi_{B,+}(\omega) = \chi_0 \frac{ \omega_0}{\omega_0 - \omega - i/T_2}.
    \label{B+}
\end{equation}
\newline
It is noteworthy that the foregoing derivation assumes a positive-valued static induction field, enabling LCP light to drive spin transitions, and resulting in the observed magnetization
\begin{align}
    \begin{split}
        & M_x = (\chi ' \cos (\omega t) + \chi'' \sin (\omega t))B_1,\\
        & M_y = (-\chi '' \cos (\omega t) + \chi' \sin(\omega t))B_1,  \\
        & M_z = \chi_0 B_0{,}
    \end{split}
\end{align}
which is neatly summarized by the magnetic susceptibility tensor
\begin{equation}
    \boldsymbol{\chi}_{M,+} = \frac{1}{2} \begin{pmatrix}
        \chi_{B,+} & -i\chi_{B,+} & 0 \\
        i \chi_{B,+} & \chi_{B,+} & 0 \\
        0 & 0 & 2\chi_0 
    \end{pmatrix},
    \label{chi+}
\end{equation}
{where the third diagonal element highlights the distinct magnetic response along the z-axis.} The time-harmonic response of the magnetic induction within the magnetized medium is then obtained from the summation over the vacuum and spin contributions, $\mathbf{B} = \mu_0(1+4\pi\boldsymbol{\chi}_{M,+})H_1 = \mu_0\boldsymbol{\mu}_{M,+}H_1$. 

Tensor $\chi_{M,+}$ in Eq.~\eqref{chi+} represents the optical response for {transitions that are induced by LCP light}. This polarization state predominantly influences the optical response near resonance frequencies in the presence of positive magnetic fields. To construct a comprehensive tensor encompassing responses for all polarization states, it is essential to augment the response of right-handed circularly polarized (RCP) light.\footnote{Note that this step was omitted in the ad-hoc model function approach in Ref.~\onlinecite{SiCpaper} leading to unphysical behaviors of the resulting tensor at low frequencies where the off-diagonal components do not vanish but should.} This augmentation is further based on the principle that any polarization state {and its response from a given medium} can be represented as a superposition of LCP and RCP components. Considering RCP light as the time-reversed analog of LCP light, the response for RCP light can be derived by applying the time-reversal operator to $\chi_{+}$.\cite{jackson_classical_1999, timereversalofemwaves} In the frequency-domain, the time reversal results in the following transformations\cite{Sigwarth2022-yg}
\begin{align}
    \begin{split}
        B_0 &\rightarrow - B_0,\\
        M_0 &\rightarrow - M_0,\\ 
        \chi_0&\rightarrow \chi_0,\\ 
         \omega_0 &\rightarrow - \omega_0. 
    \end{split}
\end{align}
This means that Eq.~\eqref{B+} transforms into 
\begin{equation}
    \chi_{B,-}= \chi_0 \frac{ \omega_0}{\omega_0 + \omega + i/T_2},
    \label{B-}
\end{equation}
and Eq.~\eqref{chi+} into 
\begin{equation}
    \boldsymbol{\chi}_{M,-} = \frac{1}{2} \begin{pmatrix}
        \chi_{B,-} & i\chi_{B,-} & 0 \\
        -i\chi_{B,-} & \chi_{B,-}& 0 \\
       0 & 0 & 2\chi_0 
    \end{pmatrix}{,}
    \label{chi-}
\end{equation}
when considering the response of RCP light as opposed to LCP light. 

The magnetic susceptibility tensor $\boldsymbol{\chi}_M$ which renders the optical response for all {magnetic field directions} is the sum of $\boldsymbol{\chi}_{M,+}$'s and $\boldsymbol{\chi}_{M,-}$'s {transverse components}
\begin{equation}
        \boldsymbol{\chi}_M =
        \frac{1}{2} \begin{pmatrix}
        \chi_{B,+}+\chi_{B,-} & -i\left[\chi_{B,+}-\chi_{B,-}\right] & 0 \\ 
        i\left[\chi_{B,+}-\chi_{B,-}\right] &  \chi_{B,+}+\chi_{B,-} & 0 \\
        0 & 0 & 2\chi_0
    \end{pmatrix}.
    \label{chitotale}
\end{equation}
With Eq.~\eqref{B+} and Eq.~\eqref{B-} we can express the on-diagonal and off-diagonal components of $\boldsymbol{\chi}_M$ in the frequency domain
\begin{equation}
\chi_{M,xx}=\chi_{M,yy} = {2}\chi_0\frac{\omega_0^2}{\omega_0^2 - \omega^2 -2i\omega/T_2},
\label{eq:chiondiagonalomega}
\end{equation}
\begin{equation}
\chi_{M,xy}=-\chi_{M,yx} = i{2}\chi_0\frac{\omega_0 \omega}{\omega_0^2 - \omega^2 - 2i\omega/T_2}.
\label{eq:chioffdiagonalomega}
\end{equation}
Equations~\eqref{eq:chiondiagonalomega} and~\eqref{eq:chioffdiagonalomega} are central to the further discussions in this work. It is noted that the on-diagonal components are equal. {Their} frequency dependence is identical to that of a harmonically broadened Lorentzian oscillator model. This is no surprise since the Bloch equations describe the motion of a magnetic moment in a harmonic potential. The off-diagonal components differ in sign and are purely imaginary in case of infinite relaxation time $T_2$. Hence, the off-diagonal components render circular dichroism, and when $T_2 < \infty$, the off-diagonal components also describe circular birefringence. The spectral behavior between on- and off-diagonal components is almost identical if $\omega \approx \omega_0$; however, it differs distinctly when $\omega \rightarrow 0$ or $\omega \rightarrow \infty$. Specifically, the off-diagonal components vanish towards infinite wavelengths, while the on-diagonal components approach $\chi_0$, the dc magnetic susceptibility. Both {components} approach zero at {high frequencies}. A further interesting feature of this model is the fact that the on-diagonal components do not change sign with reversal of the external induction field, $B_0$, while the off-diagonal terms do. This phenomenon was noted previously during analysis attempts of THz EPR ellipsometry measurements using an ad-hoc model approach, which failed to explain the observed sign changes in Mueller matrix elements upon field reversal.\cite{SiCpaper} Finally, it can be shown that Eqs.~\ref{eq:chiondiagonalomega} and~\ref{eq:chioffdiagonalomega} always lead to loss and no gain for $0 \leq T_2$, regardless of direction of external induction $B_0$ and choice of polarization. 

To construct the full permeability tensor, we consider the external induction field parallel direction $z$, and $M_z = M_0 = \chi_0 B_0$, then
\begin{widetext}
    \begin{equation}
        \boldsymbol{\mu_M} = \boldsymbol{1} + 4\pi \boldsymbol{\chi}_M = \begin{pmatrix}
            1 + {4}\pi \frac{\chi_0\omega_0^2}{\omega_0^2 - \omega^2 -2i\omega/T_2} & -i{4}\pi\frac{\chi_0\omega_0 \omega}{\omega_0^2 - \omega^2 - 2i\omega/T_2} & 0 \\
            i{4}\pi\frac{\chi_0\omega_0 \omega}{\omega_0^2 - \omega^2 - 2i\omega/T_2} & 1 + {4}\pi \frac{\chi_0\omega_0^2}{\omega_0^2 - \omega^2 -2i\omega/T_2} & 0 \\
            0 & 0 & 1 + 4\pi \chi_0
        \end{pmatrix}.
        \label{fullmu}
    \end{equation}
\end{widetext}

\subsection{Bloch permeability eigenfunction \texorpdfstring{$\mu_{Bl}$}{} }   
The Bloch permeability $\mu_{Bl}$ is introduced as {the determinant of the permeability tensor which yields}
\begin{equation}
    \mu_{Bl} = 1 + 4\pi\chi_{M,xx} = 1 +8\pi\chi_0\frac{\omega_0^2}{\omega_0^2 - \omega^2 -2i\omega/T_2}. 
\label{eq:mubloch_introduction}
\end{equation}
Note that $\chi_{M,xx}$ does not change sign upon change of the magnetic field direction. {Therefore, $\mu_{Bl}$ does not depend on the external field direction.} Furthermore, the permeability tensor can now be written in the following form:
\begin{equation}
    \boldsymbol{\mu}_{M} = \frac{1}{2} \begin{pmatrix}
        1 + \mu_{Bl} & i\frac{\omega}{\omega_0}(1 -  \mu_{Bl}) &0 \\
        -i\frac{\omega}{\omega_0}(1 - \mu_{Bl}) & 1 + \mu_{Bl}&0 \\
        0 & 0 & \frac{1}{2}(1 + 4\pi\chi_0)
    \end{pmatrix},
\end{equation}
{and we obtain} that {$\mu_{Bl}$} is indeed {eigenfunction to the magnetic component of the electromagnetic wave} for LCP {(RCP)} light at frequencies close to resonance {$\omega_0$ (-$\omega_0$)}, which is easily confirmed by multiplying the tensor with an LCP {(RCP)} magnetic field phasor $\boldsymbol{H}_+=(1,i{,0})^T$ {($\boldsymbol{H}_-=(1,-i{,0})^T$)}

\begin{widetext}
    \begin{align}
    \begin{split}
        \boldsymbol{B} = \boldsymbol{\mu}_{M}\boldsymbol{H}_{{\pm}} = &\frac{1}{2} \begin{pmatrix}
            1 + \mu_{Bl} & i\frac{\omega}{\omega_0}(1 -  \mu_{Bl}) & 0 \\
            -i\frac{\omega}{\omega_0}(1 - \mu_{Bl}) & 1 + \mu_{Bl} & 0\\
        0 & 0 & \frac{1}{2}(1 + 4\pi\chi_0)
        \end{pmatrix} \begin{pmatrix}
            1 \\ {\pm}i \\ 0
        \end{pmatrix}{|_{\omega\rightarrow \pm\omega_0}} \approx \\ &\frac{1}{2} \begin{pmatrix}
            1 + \mu_{Bl} & {\pm}i(1 -  \mu_{Bl}) & 0 \\
            {\mp}i(1 - \mu_{Bl}) & 1 + \mu_{Bl}& 0\\
        0 & 0 & \frac{1}{2}(1 + 4\pi\chi_0)
        \end{pmatrix} \begin{pmatrix}
            1 \\ {\pm}i \\ 0
        \end{pmatrix} = \mu_{Bl} \boldsymbol{H}_{{\pm}}.
        \label{muplus}
        \end{split}
    \end{align}    
\end{widetext}
As such, considering only one circular polarization mode, the scalar function $\mu_{Bl}$ can be sufficient to represent the optical response at frequencies close to the resonant frequencies. {{$\mu_{Bl}$} is measured and modeled in this work here and will be discussed in Sect.~\ref{sec:Fe3+EPR}. Note specifically that $\mu_{Bl}$ is eigenfunction near resonance for LCP at positive field $B_0$ and eigenfunction for RCP at negative field $B_0$. At resonance, the magnetic permeability response is unity both for RCP at positive field and for LCP at negative field. At frequencies outside resonance, the behavior of the permeability is more complex, and this will be discussed below in Sect.~\ref{sec:BlochEigenpolModel}.} {Note} {that the polarization of the light is also affected by the dielectric tensor. The {calculation of the polarization state of the propagating electromagnetic waves within} the material {is discussed in Appendix}~\ref{app:GSE}.}

\subsection{Bloch model extrapolated to high spin systems}\label{sec:BlochHighSpin}
The Bloch model, originally formulated on a rate equation for a two-level system, presents limitations when applied directly to a high spin system with $2S+1$ levels. Effectively, this results in $2S$ distinct EPR active species corresponding to the $2S$ spin projections with the lowest energy. Consequently, the rate equations need to be modified if there is any interconversion between the species,\cite{boltonchapter10} by, for instance, utilizing the Bloch-McConnell equations. We still motivate using the simple Bloch model as we operate under the condition $B_1<<B_0$, resulting in minimal interconversion by photon absorption. Furthermore, we assume an absence of interactions between the constituent spins within the system as the sample is weakly doped. Effectively, this means that each spin projection is considered to be a separate spin species. Secondly, ZFS results in strongly orientation-dependent effective g-factors, which introduce inhomogeneous broadening due to small variations in the orientation of the defects. This is commonly called g-strain\cite{easyspin, PILBROW1984186} and results in underestimated relaxation times.

Subsequently, for a spin $S>1/2$ system, the frequency-dependent Bloch susceptibility is extrapolated from the 2-state model under the assumption that the $2S$ spin transitions adhere to the line shape predicted by the derived Bloch susceptibility, i.e.,
\begin{gather}
    \begin{split}
        \boldsymbol{\chi}_M^{xy} = \sum_{j=1}^{2S}  
        \frac{\chi_{0,j}\omega_{0,j}}{\omega_{0,j}^2 - \omega^2 -2i\omega/T_{2,j}}\begin{pmatrix}
            \omega_{0,j} & -i\omega \\ i\omega & \omega_{0,j}
        \end{pmatrix},
    \end{split}
\end{gather}
where the superscript $^{xy}$ denotes only the response in the xy-plane, or equivalently
\begin{widetext}
\begin{gather}
    \begin{split}\label{eq:highspinsumdecomposition}
        \boldsymbol{\chi}_M^{xy} = \sum_{j=1}^{2S}  
        \frac{\chi_{0,j}\omega_{0,j}}{\omega^2_{0,j} - \omega^2 -2i\omega/T_{2,j}}\left[\omega\begin{pmatrix}
            1 & -i \\ i & 1
        \end{pmatrix}+(\omega_{0,j}-\omega)\begin{pmatrix}
            1 & 0 \\ 0 & 0
        \end{pmatrix}
        +
            (\omega_{0,j}-\omega)\begin{pmatrix}
            0 & 0 \\ 0 & 1
        \end{pmatrix}\right].
    \end{split}
\end{gather}
\end{widetext}
where $\chi_{0,j}$, $\omega_{0,j}$ and $T_{2,j}$ denote the contribution to the {dc} magnetic susceptibility, the resonance frequency, and the transverse relaxation time for a given spin projection {$j = 1,\dots2S$}, respectively. Here $j=1$ corresponds to the transition from the lowest to the second lowest-lying state, $j=2$ from the second to the third lowest-lying state, etc.

\subsection{Bloch eigenmagnetic polarizability model}\label{sec:BlochEigenpolModel}
Inspired by Max Born's description of lattice dynamics in crystalline materials,\cite{BornHuang1954} an eigendielectric displacement vector dyad summation was recently proposed as a physical model approach to render the measured dielectric function tensor across the spectral range of $j=1,..,N$ long-wavelength active phonon modes\cite{2016SchubertInvariant,PhysRevB.93.125209,PhysRevB.95.165202,PhysRevB.97.165203,PhysRevB.99.184302,10.1063/1.5135016}
\begin{equation}\label{eq:epssum}
\varepsilon=\varepsilon_\infty+\sum^{N}_{j=1}\varrho_{j}(\mathbf{\hat{e}}^{\dagger}_{j}\otimes\mathbf{\hat{e}}_{j}),
\end{equation}
where $\varrho_l$ is a complex-valued response function representing dispersion and loss caused by phonon mode $l$, $\mathbf{\hat{e}}_j$ is a vector whose direction renders the maximum response of phonon mode $j$ to the electric field component of an electromagnetic wave, $\dagger$ indicates transpose and complex-conjugate, and $\otimes$ is the dyadic product. $\mathbf{\hat{e}}_j$ further defines a normal to the plane within which the electric field cannot excite mode $j$. The obvious advantage of this summation approach lies in the possibility to add contributions of individual dipolar linear optical excitations to the total response function of a given material, i.e., the dielectric function tensor. The eigenpolarization direction parameter $\mathbf{\hat{e}}_j$ and eigendielectric function $\varrho_j$ were demonstrated as necessities to correctly describe and determine phonon mode frequency parameters including their amplitudes and broadening information for materials with low-symmetry lattice structures, i.e., monoclinic and triclinic crystal systems.\cite{2016SchubertInvariant,PhysRevB.93.125209,PhysRevB.95.165202,PhysRevB.97.165203,PhysRevB.99.184302,10.1063/1.5135016} The approach is generally valid for all symmetries, and correctly predicts generalized, coordinate-invariant representations of the dielectric functions and coordinate-invariant formulations of the Lyddane-Sachs-Teller relationship,\cite{2016SchubertInvariant} the evolution of longitudinal phonon mode coupling with free charge carriers,\cite{10.1063/1.5089145} the existence of hyperbolic shear polaritons,\cite{Passler2022} and the anisotropic properties of band-to-band transitions in low-symmetry solid-state materials,\cite{PhysRevB.96.245205} for example.

It is of interest here to analyze the Bloch susceptibility model for its representation in terms of an analogous summation and the aim is to identify a concept analogous to the summation of lattice excitations, where the occurrence of multiple magnetic excitations can be added arbitrarily with respect to their magnetic polarization properties. The question that follows is, can a similar eigenvector be found which represents a magnetic excitation and a response function be rendered that contains dispersion and loss. Inspecting the result in Eq.~\ref{eq:highspinsumdecomposition} one recognizes a dyad decomposition into three contributions such that
\begin{equation}\label{eq:musum1}
\mu=1+\sum^{2S}_{j=1}\left(\varrho^{CP}_{j}(\mathbf{\hat{a}}^{\dagger}_{j}\otimes\mathbf{\hat{a}}_{j})+\varrho^{LP}_{j}\left[\mathbf{\hat{x}}^{\dagger}_{j}\otimes\mathbf{\hat{x}}_{j}+\mathbf{\hat{y}}^{\dagger}_{j}\otimes\mathbf{\hat{y}}_{j}\right]\right),
\end{equation}
where $\mathbf{\hat{a}}_j=(1,-i,0)$, $\mathbf{\hat{x}}_j=(1,0,0)$, and $\mathbf{\hat{y}}_j=(0,1,0)$. Eq.~\ref{eq:musum1} is the Bloch eigenmagnetic polarizability model and is a central finding of this work. We note with interest that dyad $(1,\pm i,0)^{\dagger}_j\otimes(1,\pm i,0)$ was shown in Ref.~\onlinecite{SiCpaper} to represent the magnetic susceptibility tensor for a purely left/right-circularly polarized eigenprocess. Eigenvectors $\mathbf{\hat{x}}_j=(1,0,0)$ and $\mathbf{\hat{y}}_j=(0,1,0)$ are the previously described linear eigenpolarization vectors.\cite{2016SchubertInvariant,PhysRevB.93.125209,PhysRevB.95.165202,PhysRevB.97.165203,PhysRevB.99.184302,10.1063/1.5135016} Note that Eq.~\ref{eq:musum1} remains unchanged under rotation around vector $\mathbf{\hat{a}}^{{\star}}_{j}\times\mathbf{\hat{a}}_{j}=(0,0,-2i)$, where $\times$ indicates the cross or vector product {and $^{\star}$ is the complex-conjugate}. Likewise, rotation around direction $\mathbf{\hat{x}}^{{\star}}\times\mathbf{\hat{y}}=(0,0,1)$ leaves Eq.~\ref{eq:musum1} unchanged. Here, this direction is parallel to the external magnetic field direction, $\frac{\mathbf{B}_0}{B_0}$. Thereby, we can also express the gyration vector of the $j$-th eigenmagnetic polarizability contribution
\begin{equation}\label{eq:gyrationvector}
g_j=\frac{i}{2}\mathbf{\hat{a}}^{{\star}}_{j}\times\mathbf{\hat{a}}_{j}=\mathbf{\hat{x}}^{{\star}}\times\mathbf{\hat{y}}.
\end{equation}
Note further that rotation of the magnetic susceptibility tensor $\mu$ relative to a given sample and ellipsometer coordinate system can be perform simply by using Euler angle rotations and rotation matrix $A$ defined in Appendix~\ref{app:Eulerrotations}. Equation~\ref{eq:musum1} is then rewritten for new $\mu'=A\mu A^{-1}$ by replacing $\hat{\mathbf{e}}_j'=\hat{\mathbf{e}}_jA^{-1}$, $\hat{\mathbf{x}}_j'=\hat{\mathbf{x}}_jA^{-1}$, and $\hat{\mathbf{y}}_j'=\hat{\mathbf{y}}_jA^{-1}$, hence, $g_j'=Ag_j$.\footnote{Note that $\mathbf{v}A^{-1}=(A^{-1})^{T}\mathbf{v}=A\mathbf{v}$, when $A$ is a unitary matrix, i.e., $AA^{-1}=AA^{T}=I$ and $I$ is the unit matrix and $\mathbf{v}$ a vector.} It is commonly assumed that gyration vectors of all $j$ spin resonances are parallel to the external magnetic field, $g_j'=\frac{\mathbf{B_0}}{B_0}$. However, this latter statement may not necessarily be true. Investigations using THz EPR ellipsometry at multiple angles of incidence, for example, could be used to test this assumption in the future.

{When the eigenvectors $\mathbf{\hat{a}}_j$, $\mathbf{\hat{x}}_j$, and $\mathbf{\hat{y}}_j$ are all equal among the $j=1\dots 2S$ magnetic transitions, then} the associated Bloch response function components obtained here are{\footnote{When the eigenvectors are not equal, i.e., when the gyration vectors differ among the $j$ resonances, then the response functions are projected onto the permeability tensor components and require further analysis.}}

\begin{gather}
    \begin{split}\label{eq:CPandLBsusceptibility}
        \varrho^{CP}_j = \frac{\chi_{0,j}\omega_{0,j}\omega}{\omega_{0,j}^2 - \omega^2 -2i\omega/T_{2,j}}{,}  \\\varrho^{LP}_j = \frac{\chi_{0,j}\omega_{0,j}(\omega_{0,j}-\omega)}{\omega_{0,j}^2 - \omega^2 -2i\omega/T_{2,j}},
    \end{split}
\end{gather}
\begin{equation}
    \begin{split}
        &\begin{cases}
            \omega \rightarrow 0 &\textrm{then } \varrho^{CP}_j \rightarrow 0  \\
            \omega \rightarrow \omega_{0,j} &\textrm{then  } \varrho^{CP}_j \rightarrow 2i\chi_{0,j}\omega_{0,j}T_2\\
                    \omega \rightarrow \infty &\textrm{then  } \varrho^{CP}_j \rightarrow  0
        \end{cases}\\
        &\begin{cases}
            \omega \rightarrow 0 &\textrm{then  }  \varrho^{LP}_j \rightarrow \chi_{0,j} \\ 
            \omega \rightarrow \omega_{0,j} &\textrm{then  }  \varrho^{LP}_j \rightarrow 0\\
            \omega \rightarrow \infty &\textrm{then  }  \varrho^{LP}_j \rightarrow  0
        \end{cases}
    \end{split}
    \end{equation}
The two functions differ subtly in their spectral dependencies. The circularly polarized process is linear in the external magnetic induction, $B_0$, while the linearly polarized process has a component which is linear and another which is quadratic in $B_0$. The circularly polarized process disappears when $\omega$ approaches zero and infinity, while at resonance, the response function is purely imaginary and the amplitude is proportional to the magnetization {associated with the $l^{\mathrm{th}}$ spin} transition, its frequency, and transverse relaxation time. At resonance, there is no linear{ly polarized} component of the magnetization. The latter also vanishes at infinite frequencies. However, at zero frequency, the linear polarizability function reveals $\chi_{0,j}$. The linear components constitute a superposition of two equivalent linear eigenmagnetic contributions polarized within the $(x,y)$ plane in this coordinate system  and the response within this plane is isotropic. 

{It is useful to introduce the Bloch susceptibility function for magnetic transition $j$, $\chi_{B{l},j}$}
\begin{equation}\label{eq:Blochsusceptibility}
        \chi_{B{l},j}=\frac{\chi_{0,j}\omega_{0,j}\omega}{\omega_{0,j}^2 - \omega^2 -2i\omega/T_{2,j}};
\end{equation}
{and rewrite Eq.~\ref{eq:CPandLBsusceptibility}}
\begin{gather}
    \begin{split}
        \varrho^{CP}_l = \chi_{B{l},j};
        \\\varrho^{LP}_l = \chi_{B{l},j}\frac{\omega_{0,j}-\omega}{\omega}.
    \end{split}
\end{gather}
{We can then express, and plot if necessary, the scalar, complex-valued Bloch permeability eigenfunction as a sum over all Bloch susceptibility functions for all magnetic transitions $j=1,\dots,2S$} 
\begin{equation}\label{eq:Blochpermeabilityeigenfunctionsum}
        \mu_{Bl}=1+\sum_{j,1}^{2S}\chi_{B{l},j}.
\end{equation}
We propose use of Eq.~\ref{eq:musum1} and its component vectors and functions as suitable decomposition approach to analyze the measured spectral appearances of real and imaginary parts of the magnetic permeability function, for example, obtained from THz-EPR ellipsometry, or magnetooptic ellipsometry investigations. {We also propose use of Eq.~\ref{eq:Blochpermeabilityeigenfunctionsum} to present and discuss the spectral response of spin transitions if such can be considered to share a common gyration vector.} Future work will illuminate properties and usefulness of this approach.

\subsection{The Hamiltonian of the S=5/2 Fe3+ defect in GaN}\label{sec:Fe3+Hamiltonian}

Iron-doped wurtzite GaN has previously been studied extensively by several groups,\cite{Kashiwagi_2007, BARANOV1997611, maier1994} where it has been shown that the spin-Hamiltonian suggested by Bleany \textit{et al.}\cite{1954BleaneyTrenam} and altered by Geschwind\cite{Geschwind} {suffices to explain the experimental results and local symmetry of the $S=5/2$ Fe$^{3+}$ defect}
\begin{widetext}
    \begin{align}
    \begin{split}
   & \mathcal{H} = g\mu_B B_0 \textbf{S}_z \textbf{cos}\theta + \frac{1}{2}g\mu_B B_0 \textbf{sin}\theta(\textbf{S}_+ + \textbf{S}_-) + D(\textbf{S}_z^2 - \frac{1}{3}\textbf{S}(\textbf{S}+1)) \\ & -\frac{1}{180}(a - F)(35\textbf{S}_z^4 - 30\textbf{S}(\textbf{S}+1)\textbf{S}_z^2 + 25\textbf{S}_z^2 - 6\textbf{S}(\textbf{S}+1) + 3 \textbf{S}^2(\textbf{S}+1)^2) \\ & \frac{\sqrt{2}}{36}a(\textbf{S}_z(\textbf{S}_+^3e^{-i3(\psi \pm \alpha)} + \textbf{S}_-^3e^{i3(\psi \pm \alpha)}) + (\textbf{S}_+^3e^{-i3(\psi \pm \alpha)} + \textbf{S}_-^3e^{i3(\psi \pm \alpha)})\textbf{S}_z).
   \end{split}
   \label{1}
\end{align}
\end{widetext}
Here, the $g$-factor is taken to be isotropic, and $\textbf{S}$, $\textbf{S}_+$, $\textbf{S}_-$ and $\textbf{S}_z$ are the usual $S=5/2$ spin matrices. Furthermore, the {effect of} ZFS is represented in {Eq.}~\eqref{1} by $D$ and $F$, which are the axial crystal field parameters to the second and fourth order, respectively, and $a$ is the cubic crystal field parameter. Kashiwagi \textit{et al.}\cite{Kashiwagi_2007} conclusively demonstrated by applying {Eq.}~\eqref{1} to conventional X- and Q-band EPR that Fe$^{3+}$ substitution of Ga$^{3+}$ in GaN occurs, and observed that there are two {non}equivalent Ga-sites in wurtzite GaN with respect to the internal crystal field {(Fig.~\ref{fig:GaN})}. The {non}equivalence is due to the two sites' different rotations with regard to the internal crystal (ligand) field, and thus have different $\alpha$-angles, as discussed {below}. In addition, the selection rule $\Delta m_s=\pm 1$, where $m_s$ is the sextet of projections, suggests that five spin transitions are allowed for each of the two sites, resulting in up to a total of ten allowed spin transitions. The angles $\theta$, $\gamma$, and $\alpha$ correspond to the angle between the magnetic field and the [0001] axis, the angle between the magnetic field projection onto (11$\Bar{2}$3) and the [10$\Bar{1}$0]-direction, and the angle between the cubic crystal field axis $a_{1,2}$ and the [$2\Bar{1}\Bar{1}0$]-direction, respectively, are depicted in Fig.~\ref{fig:GaN}. For all measurements, the sample was rotated such that it rendered the two distinct sites equivalent in terms of the crystal field, consequently resulting in a total of five discreet spin transitions observable within each spectrum. This is the case when the magnetic field is aligned with {$\langle$1$\Bar{1}$00$\rangle$. Consequently, the spin species corresponding to each of the two sites are treated as one, which is an exception for this given rotation.

\begin{figure}[!tbp]
  \begin{center}
 \includegraphics[width=1\linewidth]{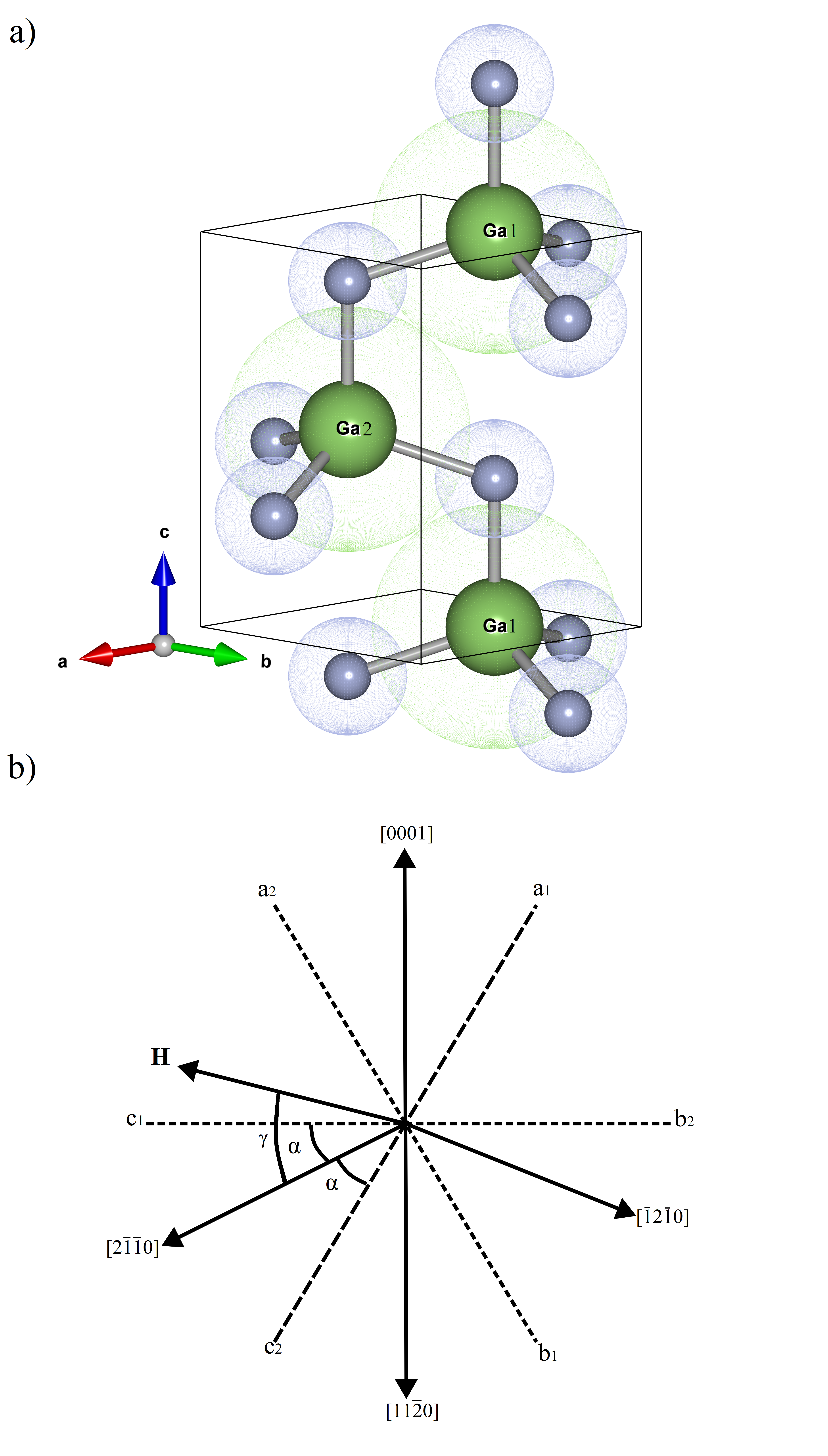}
    \caption{a) Wurtzite crystal structure of GaN along the [000$\bar{1}$] direction (c-axis). The Nitrogen atoms are depicted in blue color and the Gallium atoms are depicted in green color and denoted by either Ga1 or Ga2, depending on the relative rotation of the two nonequivalent Gallium sites with respect to the internal ligand field. Produced with computer software VESTA. \cite{VESTA} b) Projection onto the {(}11$\Bar{2}$3{)} plane of the crystalline field-axes of the two Gallium sites, as given by Ref. \onlinecite{Geschwind}. {The definitions of the angles $\gamma$ and $\alpha$ are schematically drawn, as well as the magnetic field $H$ and a few relevant crystallographic directions.}}
    \label{fig:GaN}
  \end{center}
\end{figure}

\subsection{The Brillouin magnetization of the S=5/2 Fe3+ defect in GaN}\label{sec:BrillouinMagGaN}

In extension to using Eq.~\eqref{fullmu} and to best-match model the frequency-dependent permeability function we augment additional constraints for our specific $S=5/2$ system. The constraints affect the amplitude values of the spin transitions at a fixed magnetic field, and are obtained from the summation of magnetic moments according to their thermal population. This constraint then permits the use of the spin volume concentration {$n_e$} as the input parameter for the MMF approach. A similar analysis was performed by Maryasov and Bowman on a $S=5/2$ system.\cite{2010BlochS12} We start by writing the {dc} magnetization as 
\begin{align}
    \begin{split}   
        & M_{0} = \frac{g\mu_B}{2} ( ( n_{+1/2} - n_{-1/2}) + 3 (n_{+3/2} - n_{-3/2}) \\& + 5 ( n_{+5/2} - n_{-5/2}) ). 
    \end{split}
    \label{m0}
\end{align}
The expression, sometimes referred to as Brillouin magnetization, can be comprehended as an outcome wherein each unpaired electron contributes $gm_s\mu_{Bl}$ to the overall magnetization, and $n_{+1/2}, n_{-1/2}$, etc., refer to the spin population volume density in level +1/2, -1/2, etc. Furthermore, Eq.~\eqref{m0} can be restructured as 
\begin{align}
    \begin{split}
         & M_{0} = \frac{g\mu_B}{2} (  5(n_{+5/2} - n_{+3/2}) + 8(n_{+3/2} - n_{+1/2}) \\ & + 9 (n_{+1/2} - n_{-1/2}) + 8(n_{-1/2} - n_{-3/2}) \\ & + 5(n_{-3/2} - n_{-5/2})) = \sum_j^5 M_{0, j}, 
    \end{split}
    \label{restruc}
\end{align}
i.e., as a sum corresponding to the five allowed spin transitions. The occupancy of each state is given by its Boltzmann factor
\begin{equation}\label{eq:BoltzmannMagnetizations}
    n_n - n_{n-1} = \frac{e^{-E_n/kT} - e^{-E_{n-1}/kT}} {\sum_j e^{-E_j/kT}} n_e.
\end{equation}
where the energy levels $E_n$ are given by the eigenvalues of {Eq.}~\eqref{1}. In the same manner, the {dc} magnetic susceptibility function is modeled as a sum of all allowed spin transitions contributing independently
\begin{equation}
	\chi_0 = \frac{ M_0}{B_0} = \sum_j^5 \frac{ M_{0, j}}{ B_0} = \sum_j^5 \chi_{0, j}.
 \label{FeChi}
\end{equation}
Hence, Eq.~\ref{eq:BoltzmannMagnetizations} provides specific ratios between amplitude strengths among the five transitions of the S=5/2 spin system for any given magnetic field $B_0$, and the scaling factor on all amplitudes provides the spin density. Because the magnetic field changes the eigenenergies, the ratios of amplitudes change as well, which can be measured by THz-EPR ellipsometry. Hence, measurements at different magnetic fields {and/or different temperatures} combined provide sensitivity to the spin density.

\section{Experiments and Methods}

\subsection{Sample}\label{sec:sample}
We conduct a quantitative assessment of the THz magnetic permeability properties of a wurtzite-structure single crystalline GaN {substrate} doped with Fe. The GaN {substrate with (0001) $c$-plane surface orientation is fabricated by hydride vapor phase epitaxy.\cite{Paskova_APL2006}The substrate thickness of 0.365~mm, as measured by THz ellipsometry, agrees very well with the value of 0.350$\pm$0.015~mm, specified by the provider Suzhou Nanowin Science and Technology Co. Ltd.} 
The lateral dimensions of the GaN sample are 20$\times$20~mm$^{2}$. The dislocation density in the GaN samples was estimated from x-ray diffraction\cite{Xie_JAP2014} and cathodoluminescence panchromatic imaging\cite{DARAKCHIEVA2008959} to be in the order of 1$\times$10$^7$cm$^{-2}$. {The GaN substrate is semi-insulating with a resistivity $>10^6$~$\Omega$.cm at 300~K.}

\subsection{THz-EPR ellipsometry}\label{Method:EPR}
Measurements of the Mueller matrix elements are performed using an in-house built THz ellipsometer system {at the Terahertz Materials Analysis Center at Lund University}. {The instrument uses a dual-rotating waveplate setup with fast continuing frequency-sweeping, {which differs from the optical Hall effect and THz EPR generalized ellipsometry instruments described in our previous works.\cite{KnightRSI2020,2018LuEllipsometer,doi:10.1063/1.4889920,SiCpaper}}} {Note that the instrument incorporates additional anisotropic polarizing optical elements allowing} for the measurements of the 15 normalized Mueller matrix elements as reported here. {It further} uses a {fixed} linear polarizer, rotating waveplate, sample, rotating waveplate, and {fixed} linear polarizer (analyzer) configuration. The instrument calibration and operation schemes for data acquisition follow the same procedures as described by Ruder~\textit{et al.}\cite{Ruder:21,Ruder:20} The anisotropic waveplates consist here of {3D}-printed plastic slanted columnar thin films. Such structures produce sufficient anisotropy in the THz spectral range to modulate the Stokes vector components at normal incidence transmission upon rotating the waveplates around its surface normal.\cite{10.1063/1.3626846,HofmannChapter11} A more detailed description of the instrument will be provided elsewhere. For measurements with the sample immersed in a magnetic field, a superconducting split-coil magnet is employed capable of creating magnetic fields from -8~T to 8~T with a field homogeneity of approximately 3000~ppm across {a central cylindric volume with a diameter of 10 mm} (Cryogenics Ltd. London UK). Further details of the magnet setup are given in Refs.~\onlinecite{SiCpaper,2018LuEllipsometer}.

{THz EPR ellipsometry} measurements were performed in the spectral range from 199-208~GHz in steps of 9.8~MHz. {The} source bandwidth {was} approximately 50~kHz. The instrument uses a solid-state synthesizer source with digital control over frequency and duty cycle.\cite{SiCpaper} The Mueller matrix elements are obtained from a subsequent collection of intensity readings at the {solid-state} detector for various settings of {polarizer, waveplate 1, waveplate 2, and analyzer as described by Ruder~\textit{et al.}\cite{Ruder:21,Ruder:20}} {The instrument then records} the full 4$\times$4 Mueller matrix{, i.e., 16 elements normalized by element $M_{11}$}. Measurements are performed in reflection configuration with the sample positioned between the split coils at 45$^{\circ}$ angle of incidence. In this configuration, the magnetic field direction aligns parallel with the incident beam. Thereby the magnetic field is oriented at an angle of 45$^{\circ}$ towards the crystallographic axis $\mathbf{c}$ of the GaN sample. Additionally, the sample is rotated around its normal (azimuthal rotation) to achieve positioning of the two nonequivalent gallium sites with respect to the static magnetic field such that their {respective} magnetic spin eigenenergy values coincide. {This is the case when the magnetic field is aligned with [1$\Bar{1}$00] or an equivalent direction as described in Sect.~\ref{sec:Fe3+Hamiltonian}.} As a result, the two quintuplets merge and appear as exactly one quintuplet in the THz-EPR ellipsometry spectra. {This orientation is selected here for the purpose of simplifying the analysis by reducing necessary model calculations because the thermodynamic distribution across one common spin system must be considered only. The purpose of the experiment is to demonstrate the approach.} The sample {temperature} is held constant during measurements. {Data were measured at temperatures of 20~K and 30~K, and} at a magnetic field strength of -7.23~T and 7.23~T. Data were also measured at zero field, and the zero-field data were subtracted from the field data in order to determine small-signal difference data which {were} then used as target data during the subsequent Hamiltonian and best-match model calculation analyses. Finally, the data collected from the -7.23 T scan was subtracted from the +7.23 T scan, to further amplify the signal strength.

\subsection{Ellipsometry model calculations}\label{sec:MMFparameteranalysis}
The ellipsometry data model analysis is performed using the Berreman\cite{Berreman:72} $4\times4$ transfer matrix approach augmented with modifications described by Schubert.\cite{PhysRevB.53.4265} The approach is briefly summarized in Appendix~\ref{app:4x4}. {A three-phase substrate-film-ambient model is used.} The ambient is air. The substrate constitutes a metallic substrate, assumed to exhibit complete reflectance to the incident electromagnetic waves. The substrate is non-magnetic and modeled as a highly conductive dielectric material using the isotropic Drude model, $\varepsilon=1-\frac{\omega^2_p}{\omega(\omega-i\gamma_p)}$ with plasma frequency $\omega_p=10^{-6}$cm$^{-1}$ and plasma scattering time $\gamma_p=0$~cm$^{-1}$. The subsequent layer is GaN with a thickness of 365~{$\upmu$m}. The dielectric properties of the GaN layer were calculated using the anisotropic dc dielectric constant values reported by Hibberd \textit{et al.},\cite{GaNrefindex} where $\varepsilon_a$ and $\varepsilon_c$ refer to the dielectric permittivity for polarization along crystal axes $a$ and $c$, respectively,
\begin{equation}
    \epsilon = 
    \begin{pmatrix}
        \varepsilon_a & 0 & 0 \\ 0 & \varepsilon_a & 0 \\ 0 & 0 & \varepsilon_c
    \end{pmatrix} 
    = 
    \begin{pmatrix}
        9.22 & 0 & 0 \\ 0 & 9.22 & 0 \\ 0 & 0 & 10.32
    \end{pmatrix},
\end{equation}
{and} wherein the coordinate system {$(x',y',z')$} of the dielectric tensor is congruent with the laboratory reference frame utilized for the ellipsometric data evaluation. {Note that due to the fact that Fe acts as compensating dopant in GaN, the material is electrically fully insulating and no Drude model contribution to the model dielectric tensor is necessary.} The magnetic permeability tensor was modeled according to Eq.~\eqref{fullmu} and the parameters were inferred from the {best-match parameter} model described below. This tensor undergoes a rotation, explained in Appendix~\ref{app:Eulerrotations}, employing Euler angles $\psi=-90^\circ$ and $\theta=45^\circ$. This rotation is imperative to align the tensor's {gyration (Eq.~\ref{eq:gyrationvector} in Sect.~\ref{sec:BlochEigenpolModel})} axis with the externally applied magnetic field and to align the $(x,z)$-plane with the incidence plane of the electromagnetic wave. Subsequently, both the dielectric and permeability tensors are integrated into the $\Hat{\Delta}$-matrix (Appendix~\ref{app:4x4}). This integration facilitates the computation of the optical response via the 4$\times$4-matrix {approach}. 

From the 4$\times$4-matrix method, the calculated Mueller-matrix elements $M_{23}$, $M_{32}$, $M_{14}$ and $M_{41}$ are compared to the best-match model. {These specific elements are most sensitive to spin transitions and are thus chosen.} A more detailed explanation of how these are calculated is presented in Appendix~\ref{app:MM}. To improve the signal-to-noise ratio, we add $M_{32}$ to $M_{23}$ and subtract $M_{41}$ from $M_{14}$, {because {it follows} from the magnetic permeability tensor model \eqref{fullmu} used in our analysis it follows that} $M_{32}=M_{23}$ and $M_{14}=-M_{41}$. {This can be seen directly in the full Mueller matrix element spectra set in the supplementary material.} As such, 'effective' Mueller-matrix elements {are calculated from experimental and model calculated data}
\begin{equation}
    M_{23{+32}} = \frac{M23 + M32}{2},
\end{equation}
and
\begin{equation}
    M_{14{-41}} = \frac{M14 - M41}{2}.
\end{equation}

\subsection{{MMF parameter regression analysis}}
\label{sec:MMFparameteranalysis2}
The best-match model was calculated using the {\textit{curve\_fit}} function from the SciPy library.\cite{2020SciPy-NMeth} The {\textit{curve\_fit}} function uses a trust region reflective algorithm, which aims to minimize the sum of squared residuals
\begin{equation}\label{eq:squaredresiduals}
    S = \sum_{i=1}^j (y_i - f_i)^2,
\end{equation}
where the sum runs over all frequencies, {$y_i=[M_{23+32},M_{14-41}](\omega_i)$ at frequency $\omega_i$, and $N$ is the total number of frequencies at which Mueller matrix elements were included into the regression analysis}. The model response $f_i$ is calculated by simultaneously optimizing parameters related to the spin-Hamiltonian $D$, $a$, $F$, $g${,} and $\theta${,} and the Bloch permeability parameters $n_e$ and $T_{2,j}$ {for all $j=1,\dots,5$ spin transitions}. Error estimates are determined as the square root of the diagonal elements of the covariance matrix, as computed by the curve-fitting procedure. {Note that no data-point uncertainty-based biasing was performed in Eq.~\ref{eq:squaredresiduals} because the experimentally determined uncertainty (variance) on every Mueller matrix element was observed to be uniform ($\approx$0.01\%, see  Mueller matrix spectra shown in supplementary material) across the spectral range investigated.}

\section{Example: spin transitions in Fe-doped GaN}\label{sec:Fe3+EPR}
{The experiment on the spin $S=5/2$ system in GaN is selected here for the purpose of demonstration. THz EPR ellipsometry was performed as described above. The Hamiltonian description of the zero-field split system is well known, and thereby is suitable for the purpose here. As outlined in earlier sections, the sample is oriented such that the two ZFS quintuplets align in their eigenvalues and thermodynamic spin distributions. As a result, only one common quintuplet is observed.  }
\subsection{Results}
\begin{figure*}[!tbp]
\centering
 \includegraphics[width=1\linewidth]{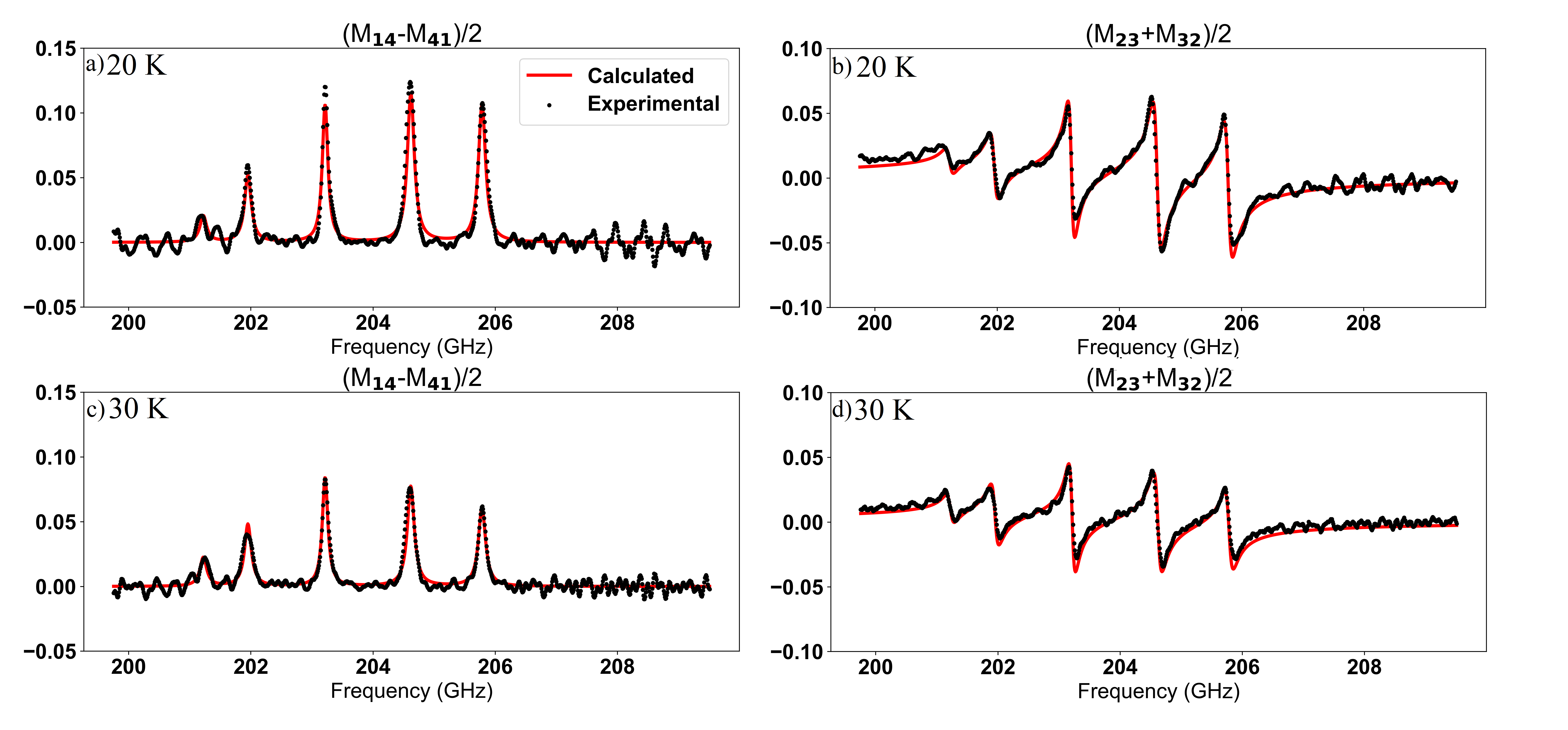}
    \caption{Results from frequency-swept THz-EPR ellipsometry measurements performed at a,b) 20~K and c,d) 30~K. The calculated best-match model is represented by the solid red line and is based on the Bloch-Brillouin model. {The black symbols indicate the experimental data.} In a,c) {$M_{14-41}=(M_{14}-M_{41})/2$} is shown, which {scales with} circular dichroism. In b,d) {$M_{23+32}=(M_{23}+M_{32})/2$} is shown, which {scales with} circular birefringence. {Experimental and model data are obtained as difference data for magnetic field $B=-7.23$~T and $B=7.23$~T.}}
    \label{fig:MMresults}
\end{figure*}

\begin{figure}[!tbp]
  \begin{center}
 \includegraphics[width=1\linewidth]{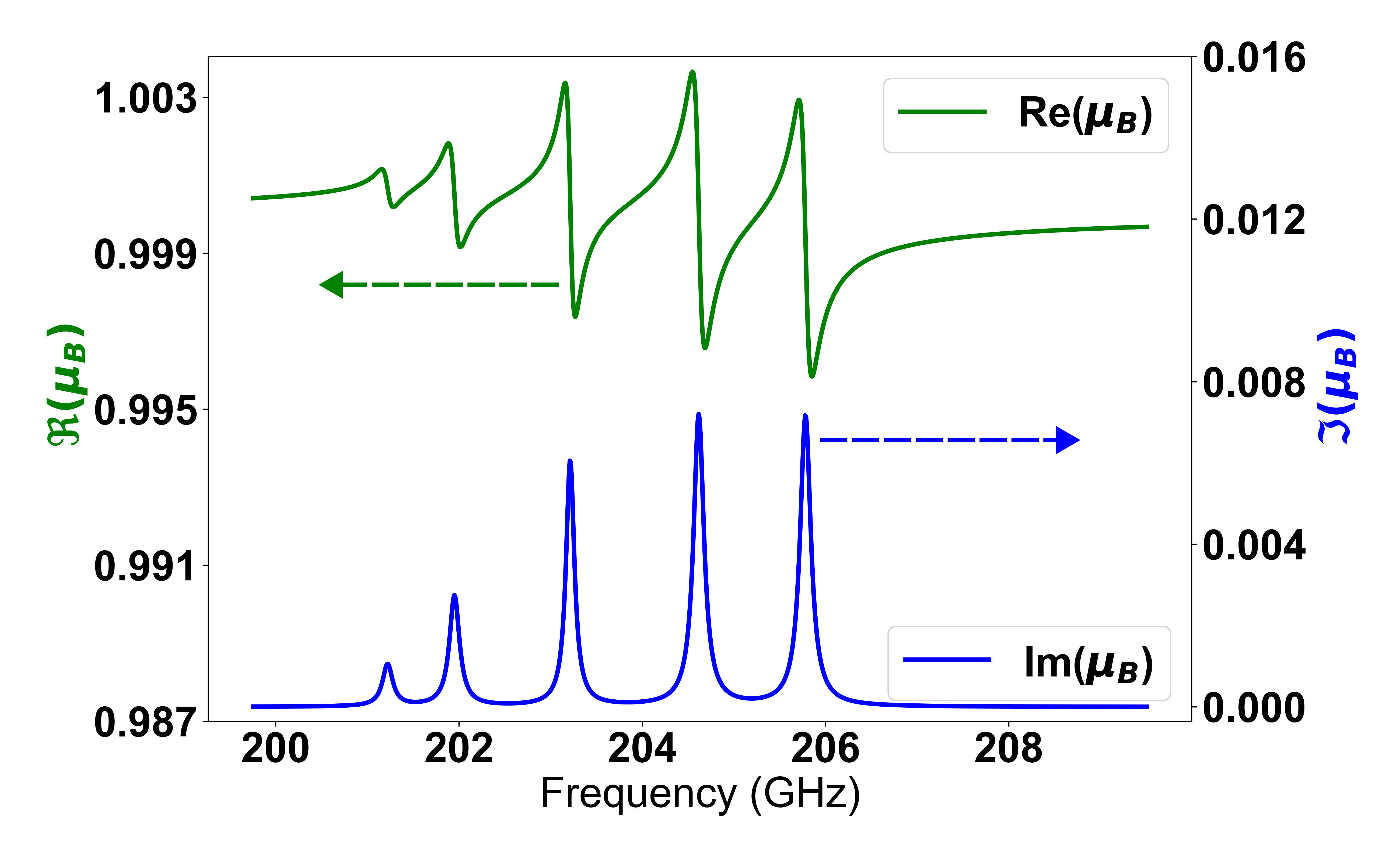}
    \caption{Extracted Bloch permeability {eigenfunction $\mu_{Bl}$} at {$B=7.23$~T and} 20~K from the best-match model {according to} \eqref{eq:Blochpermeabilityeigenfunctionsum}. The lower part shows the imaginary part of $\mu_{Bl}$ as a solid blue line, and the upper half shows the real part of $\mu_{Bl}$ as a solid green line. {The distribution of amplitudes across the five spin transitions is the result of the thermodynamic Brillouin model, the eigenfrequencies are the result of the Hamiltonian calculations predicting the ZFS spin transition in this example.}}
    \label{fig:muplus}
  \end{center}
\end{figure}

\begin{table*}
\centering
\caption{Summary of model parameters {obtained from analysis of} the THz-EPR ellipsometry {measurements} on iron-doped GaN. {The parameters were determined by minimizing the least-squares difference between the best-match model and experimental data gathered at temperatures of 20 and 30 K. The optimization process was confined to the Mueller-matrix elements $M_{23}$, $M_{32}$, $M_{14}$, and $M_{41}$.} All parameters are shown with error bars according to one standard deviation as approximated from the square roots of the covariance matrix.}
\label{tab:results}
\begin{ruledtabular}
\begin{tabular}{llllllll}
&$D$ & $a$& $F$ & $g$ & $\theta$ & $n_e$ \\
& MHz & MHz& MHz& n/a & $^{\circ}$ & cm$^{-3}$  \\
\hline
This work&-2250 $\pm$ 14 & 210 $\pm$ 280 & 15 $\pm$ 280 & 2.008200 $\pm$ 8$\cdot10^{-6}$ &  44.4 $\pm$ 0.05 & 1.92$\cdot10^{18}$ $\pm$ 1.57$\cdot10^{16}$    \\
Kashiwagi~\textit{et al.}$^a$ &-2290 &240  &-27 &2.008 &-  &- \\
Maier~\textit{et al.}$^b$ &-2138 &144  &-12 &1.995 &-  &- \\
Baranov~\textit{et al.}$^c$ &$|2144|$ &-  &- &1.994 &-  &- \\
\end{tabular}
\end{ruledtabular}
$^a$ X- and Q-band EPR, Ref.~\onlinecite{Kashiwagi_2007}.\\
$^b$ Ref.~\onlinecite{maier1994}.\\
$^c$ X-band EPR, Ref.~\onlinecite{BARANOV1997611}.
\label{tabZFSg}
\end{table*}

\begin{table}
\centering
\caption{Measured transverse relaxation time $T_{2,j}$ parameters and one standard deviation in units of $10^{-11}$s for various temperatures.}
\label{tabT2}
\begin{ruledtabular}
\begin{tabular}{lllllll}
 $T$& $T_{2,1}$    & $T_{2,2}$    & $T_{2,3}$    & $T_{2,4}$    & $T_{2,5}$   \\
  K& $10^{-11}$s  & $10^{-11}$s  & $10^{-11}$s    & $10^{-11}$s  & $10^{-11}$s \\
\hline
20  & 224$\pm$5 & 232$\pm$5 & 288$\pm$7 & 240$\pm$10  & 238$\pm$24   \\
30  & 215$\pm$5 & 217$\pm$4  & 260$\pm$5 & 191$\pm$6 & 223$\pm$13  \\
\end{tabular}
\end{ruledtabular}
\end{table}

Figs.~\ref{fig:MMresults}(a,b) {and Fig.~\ref{fig:MMresults}(c,d)} show experimental and best-match model calculated data for $M_{14{-41}}$ and $M_{23{+32}}$ at 20~K and at 30~K{, respectively. Note the different amplitude distributions between the different temperature measurements. The full sets of Mueller matrix data taken at the two magnetic fields at 20~K are shown in the supplementary material. The best-match model calculated data are obtained by varying the Hamiltonian parameters for calculation of the eigenenergies $\omega_{0,j}$, the total spin density $n_e$, and the transverse relaxation time parameters for each transition. An excellent match between experimental and model-calculated data is obtained. The match {illustrates} the correctness of the model derived in this work. {Particularly}, the match upon reversal of field, which changes signatures such that $M_{14}(B)=M_{41}(-B)$, $M_{23}(B)=-M_{32}(-B)$ as well as $M_{14}(B)=-M_{41}(B)$ and $M_{23}(B)=M_{32}(B)$ is {consistent} between experiment and theory. This observation is the first to our knowledge because no Mueller matrix elements have been reported thus far for spin resonances in the THz spectral range.} {Eq.~}\eqref{fullmu} can account for the expected sign change upon {magnetic} field reversal experimentally observed {here, which} was not the case with the ad-hoc model {used} previously {for} THz-EPR ellipsometry data analysis.\cite{SiCpaper} {There}, the permeability tensor {suffered from being} invariant to a change of the sign of $\omega_0$. 

The congruence between the model and experimental observations is {further} noteworthy {because} the amplitude variations are consistent with predictions derived from the Brillouin magnetization theory, and the spectral line shapes align with the theoretical projections based on the Bloch equations. {This can be seen by the amplitude distribution match in Figs.~\ref{fig:MMresults}(a-d) as well as in the full Mueller matrix displayed in the supplementary materials.} The {extracted} Bloch permeability {$\mu_{Bl}$} as given by {Eq.}~\eqref{eq:Blochpermeabilityeigenfunctionsum} {obtained from the experimental data at 7.23~T and} 20~K is shown in Fig.~\ref{fig:muplus}. This result highlights the ability of the method to measure {and determine} both the imaginary and real part of the frequency-dependent permeability tensor and thus yield access to further polarization properties of {a given magnetic transition} as well as the possibility of inferring {dc} properties by extrapolating the value of $\mu_{Bl}$ at $\omega=0$. 

{The ZFS, $g$-factor, and spin density parameters deduced from our best-match model calculations are cataloged in Table~\ref{tabZFSg}. The transverse relaxation times are summarized in Table~\ref{tabT2}.}

\subsection{Discussion}
The ZFS parameters deduced here agree well with those reported in prior studies.\cite{Kashiwagi_2007, BARANOV1997611, maier1994} Nevertheless, the substantial standard deviations{, in particular for parameters $a$ and $F$} indicate that multiple combinations of the ZFS parameters can produce {equally} satisfactory model {calculations matching our data obtained} for a singular sample {azimuth (rotation) orientation.} Additionally, {a small} angular offset between the magnetic field direction and the sample plane was {observed and is due to a slight misalignment of} the magnet cryostat to dampen standing waves {within the inner wedged diamond windows of the cryostat.} The quantitatively determined spin volume concentration $n_e$ is {in agreement with} the nominal range of {iron concentration of} this specific sample. In light of these findings, a {future} comprehensive analytical investigation is warranted to ascertain the reliability and accuracy of this technique for measuring spin volume concentration {by augmenting further magnetic field, temperature, and sample rotation data.} 

The calculated transverse relaxation times are presented in table~\ref{tabT2}. It is important to {emphasize} that these relaxation times are {inflated} due to the {rather large magnetic field broadening introduced by our} superconducting magnet, which is not engineered for optimal field homogeneity. This lack of flatness introduces a notable broadening of the observed transitions.{\footnote{A diagram of the field distribution within the sample plane is shown in the supplementary material in Ref.~\onlinecite{SiCpaper}}.} To obtain more {accurate} relaxation time {parameters}, a thorough consideration of the broadening {causes attributable} to both the magnetic field inhomogeneity and $g$-strain is required. However, such an in-depth analysis falls beyond the scope of this article. {It is noted that the $T_{2,j}$-values differ among the transitions}. {The cause of this could be due to several reasons, such as the two nonequivalent sites not perfectly overlapping or the width of the transitions being unequal because of the difference in how the spin projections interact with the local environment}. The longitudinal relaxation times {are not accessible from this steady-state continuous-wave single-frequency type measurement, and time-dependent measurement approaches are necessary. Such are conceivable using solid-state source- and detector-based techniques and could be the subject of further investigations.} We propose that THz-EPR ellipsometry is a robust technique for investigating transverse spin relaxation times, offering significant potential for future studies in this domain.

{The pairs of Mueller matrix elements $M_{14}$, $M_{41}$ and $M_{23}$, $M_{32}$ are routinely associated with circular dichroism and circular birefringence.\cite{Arteaga2023} Hence, one can interpret data in Figs.~\ref{fig:MMresults}(a,b) with circular dichroism and in Figs.~\ref{fig:MMresults}(c,d) with circular birefringence. The former is then expected to appear as a Lorentzian absorption line, while the latter is expected to appear as a form with increasing normal dispersion towards resonance, negative dispersion across resonance, and normal dispersion approaching zero above resonance. Note that all features shown in Figs.~\ref{fig:MMresults}(a-d) invert with magnetic field reversal, while spectra shown in Fig.~\ref{fig:muplus} remain unchanged. This is due to the definition we chose in Sec.~\ref{sec:BlochEigenpolModel}. Then, function $\mu_{Bl}$ represents the LCP circular birefringence in its real part and circular polarized loss in its imaginary part for positive field. Upon field reversal, $\mu_{Bl}$ then represents the RCP circular birefringence in its real part and circular polarized loss in its imaginary part. Both functions are unchanged between LCP at positive and RCP at negative fields. Hence, $\mu_{Bl}$ directly represents the chiroptical properties of magnetic resonance transitions. We note that signatures observed in our experiment relate to electron spin resonance. Antiferromagnetic resonance may reveal also signatures opposite to those observed here.}

Future research trajectories may consider the implementation of advanced versions of the Bloch equations to delve deeper into magnetic resonance phenomena. This includes the Bloch-Bloembergen equations\cite{BLOEMBERGEN1949386} {which} modify the external magnetic field to an effective field that encompasses the internal field generated by the sample, thereby making it more suitable for systems with higher spin volume concentrations. This modification allows for a more accurate representation in ferromagnetic resonance studies. In parallel, both the Gilbert equation\cite{gilbert} and Landau–Lifshitz equation\cite{LANDAU199251} offer a reformulation where phenomenological relaxation times are substituted with a relaxation rate that is proportional to the time derivative of the magnetization, accompanied by a damping parameter. This damping parameter is crucial in the study of energy dissipation within the system, particularly relevant for the investigation of spin waves and strong ferromagnets characterized by interacting magnetic moments and domain formation.\cite{Landau-Gilbert-Overview} The Bloch-McConnell\cite{1958McConnell} equations represent another extension of the original Bloch equations, designed to include the dynamics of chemical exchange processes in NMR studies. These modified equations are essential for examining the interaction of nuclear spins in systems where species undergo interconversion, leading to distinct NMR signatures. By integrating terms for chemical exchange rates, the Bloch-McConnell equations enhance the capability to analyze the impact of such exchanges on the system's magnetization. The Bloch equations can also be applied to pulsed magnetic resonance \cite{HOULT197969}, making it feasible to study fast processes such as spin dynamics in biological samples.\cite{pulsed_epr_bio}

\section{Conclusions}
{We presented a generalized approach based on Bloch’s equation of motion of nuclear magnetic moments. Our approach can model the frequency, magnetic field, moment density, and temperature dependencies of the magnetic permeability in magnetic resonance. Our model predicts polarization properties in electromagnetic wave interactions, which can be observed, for example, in full polarization-resolved Mueller matrix element spectra measured across magnetic resonances as a function of frequency, magnetic field, spin density, and temperature. Thermodynamic considerations are augmented using the concept of Brillouin magnetization. Hamiltonian perturbation approaches for ZFS and Zeeman interaction are incorporated further and the magnetic eigenvalue spectrum, spectral amplitude distribution, and geometrical orientation parameters of the magnetic moment density are obtainable from comparing the generalized model approach to experimental data. We demonstrate the validity of our approach by analyzing the oblique angle of incidence terahertz spectral range magnetic field ellipsometry to detect electron spin resonance quintuplet transitions in wurtzite-structure GaN doped with iron. Measurements at magnetic fields of $\pm$7.23~T and cryogenic temperatures of 20~K and 30~K detail the occurrence of linear and circular birefringence and dichroism associated with each of the zero-field split spin transitions in the $S=5/2$ system. We derive and discuss the spectral dependence of the magnetic susceptibility function and obtain the temperature and magnetic field-dependent Hamiltonian parameters and spin density. The 15 independent elements of the normalized Mueller matrix for both positive and negative magnetic fields are matched excellently with our model. We propose the employment of our approach to study magnetic resonance in ferromagnetic, antiferromagnetic, and nuclear magnetic resonance spectroscopy.

\section{Acknowledgments}
This work is supported by the Swedish Research Council under Grants No. 2016-00889 and No. 2022-04812, by the Knut and
Alice Wallenberg Foundation funded grant “Wide-bandgap semiconductors for next generation quantum components” (Grant No. 2018.0071), by the Swedish Foundation for Strategic Research under Grant No. EM16-0024, by the Swedish Governmental Agency
for Innovation Systems VINNOVA under the Competence Center Program Grant No. 2022-03139, and by the Swedish Government Strategic Research Area NanoLund and in Materials Science on Functional Materials at Link\"oping University, Faculty Grant SFO Mat LiU No. 009-00971. M.S. acknowledges support by the National Science Foundation under awards ECCS 2329940, and OIA-2044049 Emergent Quantum Materials and Technologies (EQUATE), by Air Force Office of Scientific Research under awards FA9550-19-S-0003, FA9550-21-1-0259, and FA9550-23-1-0574 DEF, and by the University of Nebraska Foundation. M.S. also acknowledge support from the J.~A.~Woollam Foundation.

\section{Appendix}

\subsection{Generalized Spectroscopic Ellipsometry}\label{app:GSE}
{The results of Mueller-matrix ellipsometry}\cite{Fujiwara} are commonly described by Mueller-Calculus, which is a mathematical framework where Mueller matrices\cite{MuellerMIT1943}
\begin{equation}
   \textbf{M}= \begin{pmatrix}
        M_{11} & M_{12} & M_{13} & M_{14} \\
        M_{21} & M_{22} & M_{23} & M_{24} \\
        M_{31} & M_{32} & M_{33} & M_{34} \\
        M_{41} & M_{42} & M_{43} & M_{44} 
    \end{pmatrix},
\end{equation}
operate on Stokes vectors\cite{Fujiwara}
\begin{equation}
    \textbf{S}=\begin{pmatrix}
        I_p + I_s \\
        I_p - I_s \\
        I_{45^\circ} - I_{-45^\circ} \\
        I_{\textrm{RCP}} - I_{\textrm{LCP}}
    \end{pmatrix},
\end{equation}
where $s,\ p,\ \pm 45^\circ $ denote linearly polarized light perpendicular, parallel, $\pm$ 45$^\circ$ to the plane of incidence, while RCP and LCP denote right and left circularly polarized light, respectively.
\newline

\subsection{4\texorpdfstring{$\times$4}{} matrix formalism}\label{app:4x4}
To compute the Mueller matrix and establish a connection between experimental data and a theoretical model, employing the 4$\times$4 formalism introduced by Berreman\cite{Berreman:72} is convenient. This formalism was further elaborated upon by Schubert.\cite{PhysRevB.53.4265} Fundamentally, the formalism is based on the transfer matrix equation, which can be written as within the coordinate system introduced in Ref.~\onlinecite{PhysRevB.53.4265}
\begin{equation}
    \begin{pmatrix}
        E_p^I \\
        E_s^I \\
        E_p^R \\
        E_s^R
    \end{pmatrix} = \textbf{L} \begin{pmatrix}
        E_p^T \\
        E_s^T \\
        E_p^B \\
        E_s^B
    \end{pmatrix},
\end{equation}
where $I$ and $R$ pertain to the electric components of the incoming and reflected electromagnetic (EM) waves, while $B$ and $T$ correspond to the electric components of the backward-traveling and transmitted portions of the outgoing EM waves. The transfer matrix $\textbf{L}$ consists of three factors:
\begin{equation}
    \textbf{L}= \textbf{L}_I^{-1} \prod_j^k \textbf{L}_j \textbf{L}_T.
\end{equation}
In this expression, the incident transfer matrix $\textbf{L}_I^{-1}$ projects the incident electromagnetic plane wave onto the sample's surface, and the exit transfer matrix $\textbf{L}_T$ projects the transmitted wave components onto the substrate. The second factor encapsulates the properties of the various layers within the sample, with $k$ representing the total number of layers. The transfer matrix for each individual layer can be expressed as:
\begin{equation}
    \textbf{L}_j = \text{exp}\left(-i \frac{\omega \hat{\Delta} d}{c}\right)
\end{equation}
where $c$ is the speed of light in vacuum, $\omega$ is the frequency of the electromagnetic wave, $d$ the thickness of the sample. The $\hat{\Delta}$-matrix is derived directly from Maxwell's equations and contains elements of the permeability and permittivity tensor in the frequency domain expansion:
\begin{widetext}
\begin{equation}
		\hat{\Delta} = 
		\begin{pmatrix}
			-q_x \frac{\varepsilon_{zx}}{\varepsilon_{zz}} &
			q_x (-\frac{\varepsilon_{zy}}{\varepsilon_{zz}}+\frac{\mu_{zx}}{\mu_{zz}}) & 
			\mu_{yx} - \frac{\mu_{yz}\mu_{zx}}{\mu_{zz}} &
			-q_x^2\frac{1}{\varepsilon_{zz}} + \mu_{yy} - \frac{\mu_{yz}\mu_{zy}}{\mu_{zz}} \\
			0 &
			-q_x \frac{\mu_{xz}}{\mu_{zz}}	&
			-\mu_{xx} + \frac{\mu_{xz}\mu_{zx}}{\mu_{zz}} &
			-\mu_{xy} + \frac{\mu_{xz}\mu_{zy}}{\mu_{zz}} \\
			-\varepsilon_{yx} + \frac{\varepsilon_{yz}\varepsilon_{zx}}{\varepsilon_{zz}} &
			 - \varepsilon_{yy} + \frac{\varepsilon_{yz}\varepsilon_{zy}}{\varepsilon_{zz}} +q_x^2\frac{1}{\mu_{zz}} &
			-q_x \frac{\mu_{zx}}{\mu_{zz}} &
			q_x (\frac{\varepsilon_{yz}}{\varepsilon_{zz}}-\frac{\mu_{zy}}{\mu_{zz}}) \\
			\varepsilon_{xx} - \frac{\varepsilon_{xz}\varepsilon_{zx}}{\varepsilon_{zz}} &
			\varepsilon_{xy} - \frac{\varepsilon_{xz}\varepsilon_{zy}}{\varepsilon_{zz}} &
			0 &
			-q_x \frac{\varepsilon_{xz}}{\varepsilon_{zz}}
		\end{pmatrix}, 
	\end{equation}
 \end{widetext}
where $q_x =n_I$sin$\theta$ is the $x$-component of the reduced wavevector $\frac{\omega}{c}\mathbf{q}=\mathbf{k}$ of the incoming wave vector $\mathbf{k}$ at the sample surface parallel to the plane of incidence, $n_I$ is the index of refraction of the isotropic incident medium, and $\theta$ is the angle of incidence to the surface normal. In this coordinate choice, the plane of incidence is the $(x,z)$ plane, the sample surface is parallel to the $(x,y)$ plane, $z$ points into the sample, and the surface is at the coordinate origin of the Cartesian system $(x,y,z)$. Given that spin transitions manifest as magnetic resonances, it becomes imperative to model the permeability tensor in addition to the permittivity tensor.
\subsection{Mueller matrix elements}\label{app:MM}
 Once the transfer matrix is calculated, determining the {Mueller} matrix elements becomes a straightforward task. For an in-depth explanation, readers are referred to references\cite{MuellerMIT1943,Fujiwara,Schubert:16, Yeh79}. To exemplify the procedure, the derivation of elements $M_{14,41}$ and $M_{23,32}$ from the result of the 4$\times$4 calculation scheme are presented here. It has been observed that the Mueller-matrix elements $M_{14,41}$ are proportional to the circular dichroism, and the Mueller-matrix elements $M_{23,32}$ are proportional to the circular birefringence caused by the spin transitions.\cite{SiCpaper} Consequently, the Mueller matrix elements $M_{23,32}$ and $M_{14,41}$ are employed for the detection of EPR signals in this study. These elements can be written in the form of transfer matrix elements $L_{k,l}$ with $k,l = 1,2,3,4$:
\begin{align}
    M_{23}= \begin{split}
        & \Re\bigg( \frac{(L_{11}L_{43} - L_{13}L_{41})(L_{11}L_{23}- L_{13}L_{21})^*}{L_{11}L_{33} - L_{13}L_{31}}  \\
        & -\frac{(L_{33}L_{21} - L_{31}L_{23})(L_{33}L_{41}- L_{31}L_{43})^*}{L_{11}L_{33} - L_{13}L_{31}}\bigg),
    \end{split}
\end{align}
\begin{align}
    M_{32}= \begin{split}
        & \Re\bigg( \frac{(L_{11}L_{43} - L_{13}L_{41})(L_{33}L_{41}- L_{31}L_{43})^*}{L_{11}L_{33} - L_{13}L_{31}}  \\
        & -\frac{(L_{33}L_{21} - L_{31}L_{23})(L_{11}L_{23}- L_{13}L_{21})^*}{L_{11}L_{33} - L_{13}L_{31}}\bigg),
    \end{split}
\end{align}
\begin{align}
    M_{14}= \begin{split}
        & \Im\bigg( \frac{(L_{11}L_{43} - L_{13}L_{41})(L_{11}L_{23}- L_{13}L_{21})^*}{L_{11}L_{33} - L_{13}L_{31}}  \\
        & -\frac{(L_{33}L_{21} - L_{31}L_{23})(L_{33}L_{41}- L_{31}L_{43})^*}{L_{11}L_{33} - L_{13}L_{31}}\bigg),
    \end{split}
\end{align}
and 
\begin{align}
    M_{41}= \begin{split}
        & \Im\bigg( \frac{(L_{11}L_{43} - L_{13}L_{41})(L_{33}L_{41}- L_{31}L_{43})^*}{L_{11}L_{33} - L_{13}L_{31}}  \\
        & -\frac{(L_{33}L_{21} - L_{31}L_{23})(L_{11}L_{23}- L_{13}L_{21})^*}{L_{11}L_{33} - L_{13}L_{31}}\bigg),
    \end{split}
\end{align}
where $*$ symbolizes complex conjugate. From these expressions it is evident that $M14$ and $M23$ are the imaginary and real part of the same function, respectively, and the same holds true for $M41$ and $M32$.

\subsection{Euler rotations}\label{app:Eulerrotations}

\begin{figure}[!tbp]
  \begin{center}
 \includegraphics[width=0.5\linewidth]{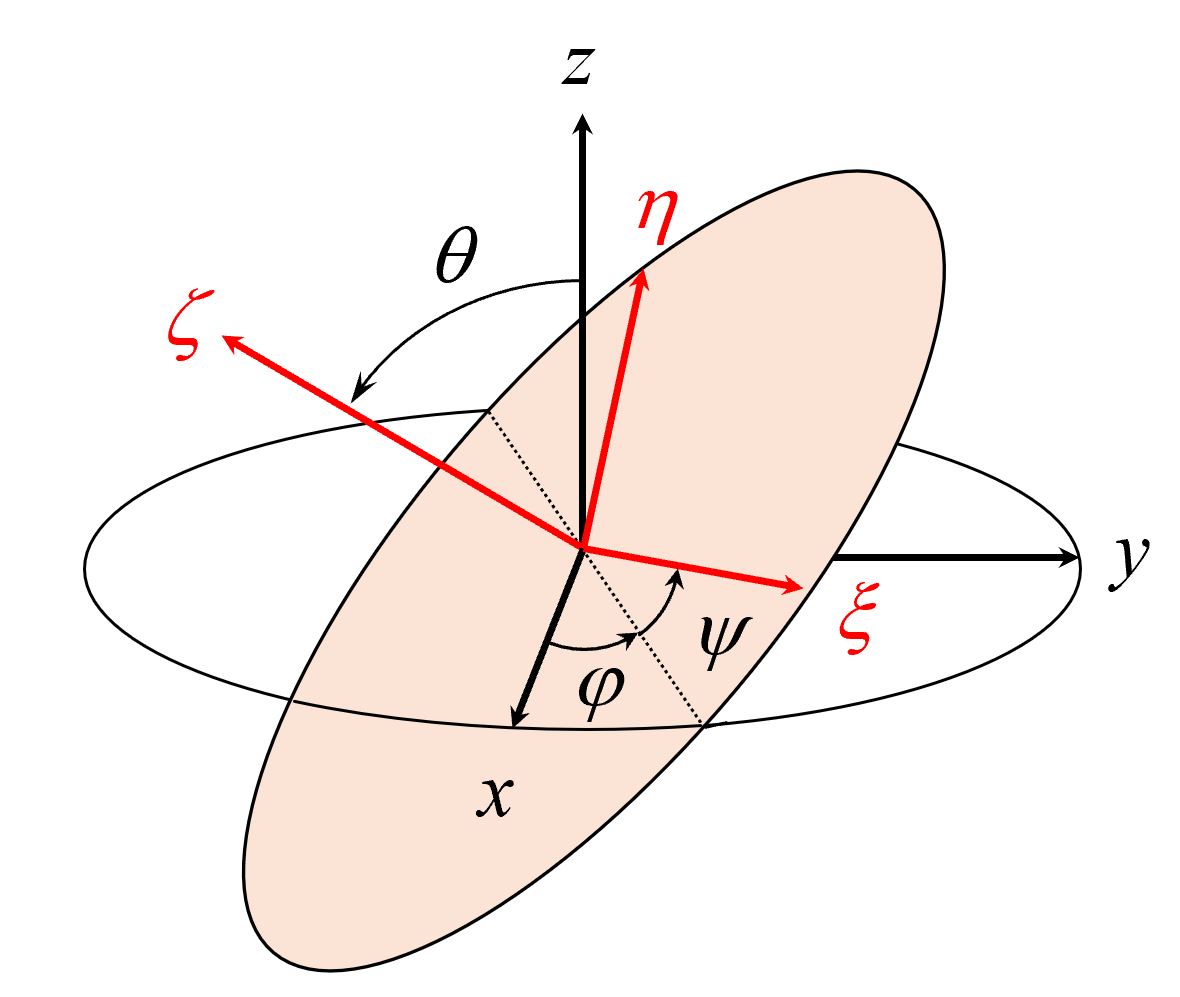}
    \caption{Definition of the Euler angles $\varphi$, $\theta$, and $\psi$ and the orthogonal rotations as provided by $\mathbf{A}$. $(\xi, \eta, \zeta )$, and $(x,y,z)$ refer to the Cartesian auxiliary and laboratory coordinate systems, respectively. Reprinted from Ref.~\onlinecite{PhysRevB.93.125209} with copyright permission by American Physical Society.}
    \label{fig:Euler}
  \end{center}
\end{figure}

For a given magnetic field orientation, Euler angle rotations are required to bring $\mu$ into the correct appearance within the ellipsometer system $(x,y,z)$. Euler rotations which perform such operations are shown in Fig.~\ref{fig:Euler}. Operation $R_1(v=\phi)$ renders rotation around $z$,  $R_2(v=\theta)$ around $x$, and $R_1(v=\psi)$ around new direction $\zeta$ with mathematically positive (negative) sense for positive (negative) arguments

\begin{equation}
R_1(v)=
\left(\begin{array}{ccc}
\cos v & -\sin v & 0 \\
\sin v &  \cos v & 0 \\
0 & 0 & 1
\end{array}\right),
\end{equation}
\begin{equation}
R_2(v)=
\left(\begin{array}{ccc}
1 & 0 & 0 \\
0 & \cos v & -\sin v \\
0 & \sin v &  \cos v
\end{array}\right).
\end{equation}

\noindent The full set of rotations, $\varphi, \theta, \psi$ indicated in Fig.~\ref{fig:Euler}, is then described by matrix $A$

\begin{equation}
A=R_1(\varphi)R_2(\theta)R_1(\psi),
\end{equation}

\noindent where $\hat{\mu}$ indicates the tensor appearance of $\mu$ in a new auxiliary system

\begin{equation}
\hat{\mu}=A\mu A^{-1}.
\end{equation}

Due to the rotational invariance of Eq.~\ref{eq:musum1} around its gyration vector in Eq.~\ref{eq:gyrationvector} rotation $R_1(\psi)$ is not needed for addressing the magnetic field direction relative to the sample system within the ellipsometer system. However, the rotational dependencies of the Hamiltonian parameters will depend on the sample orientation relative to the ellipsometer system and the magnetic field direction. For proper sample rotation to align the dielectric tensor within the ellipsometer system, all three Euler rotations maybe necessary.

\bibliography{refs}
\newpage
\begin{figure*}[htbp]
\centering
\includegraphics[page=1, width=\textwidth]{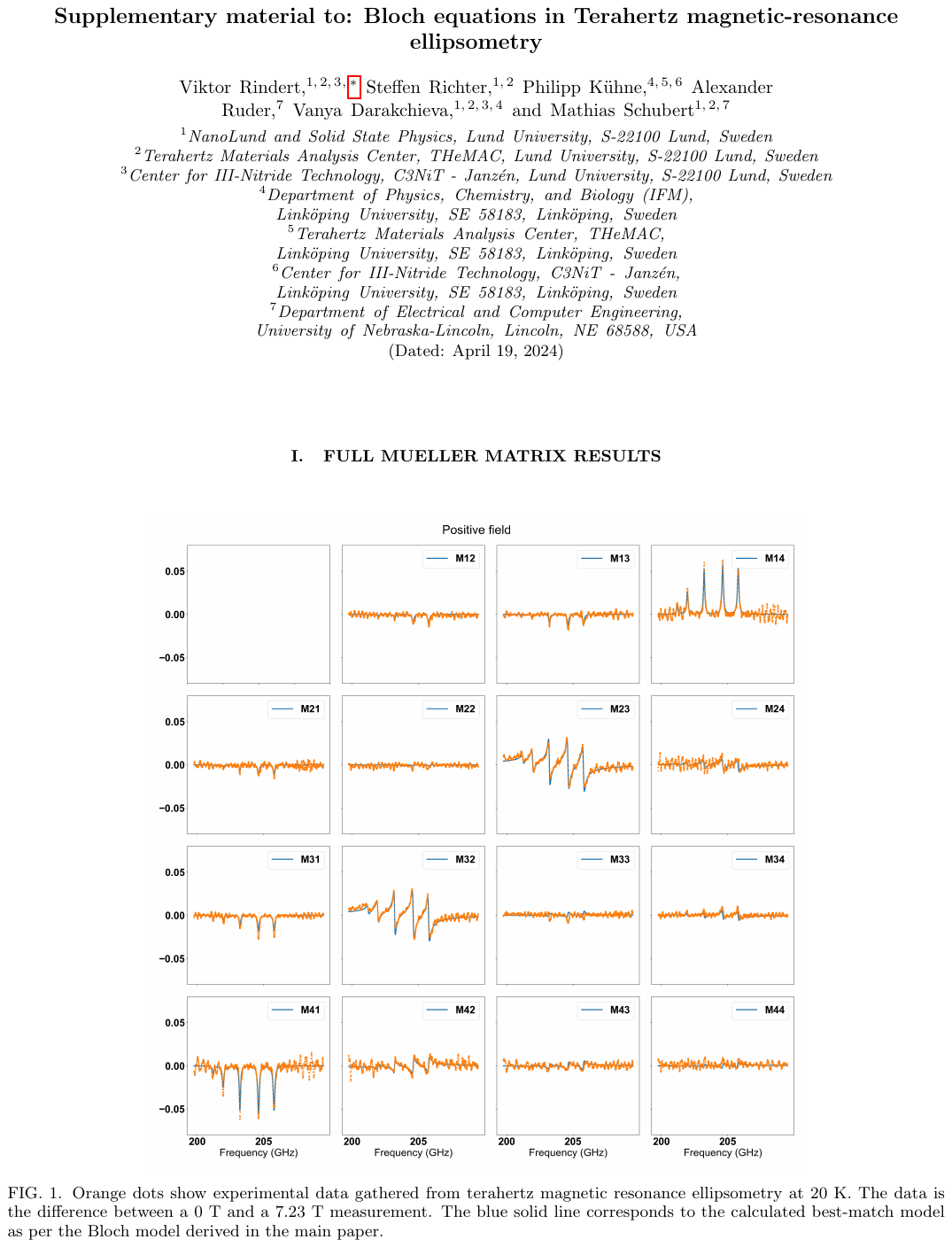}
\end{figure*}
\newpage
\begin{figure*}[htbp]
\centering
\includegraphics[page=2, width=\textwidth]{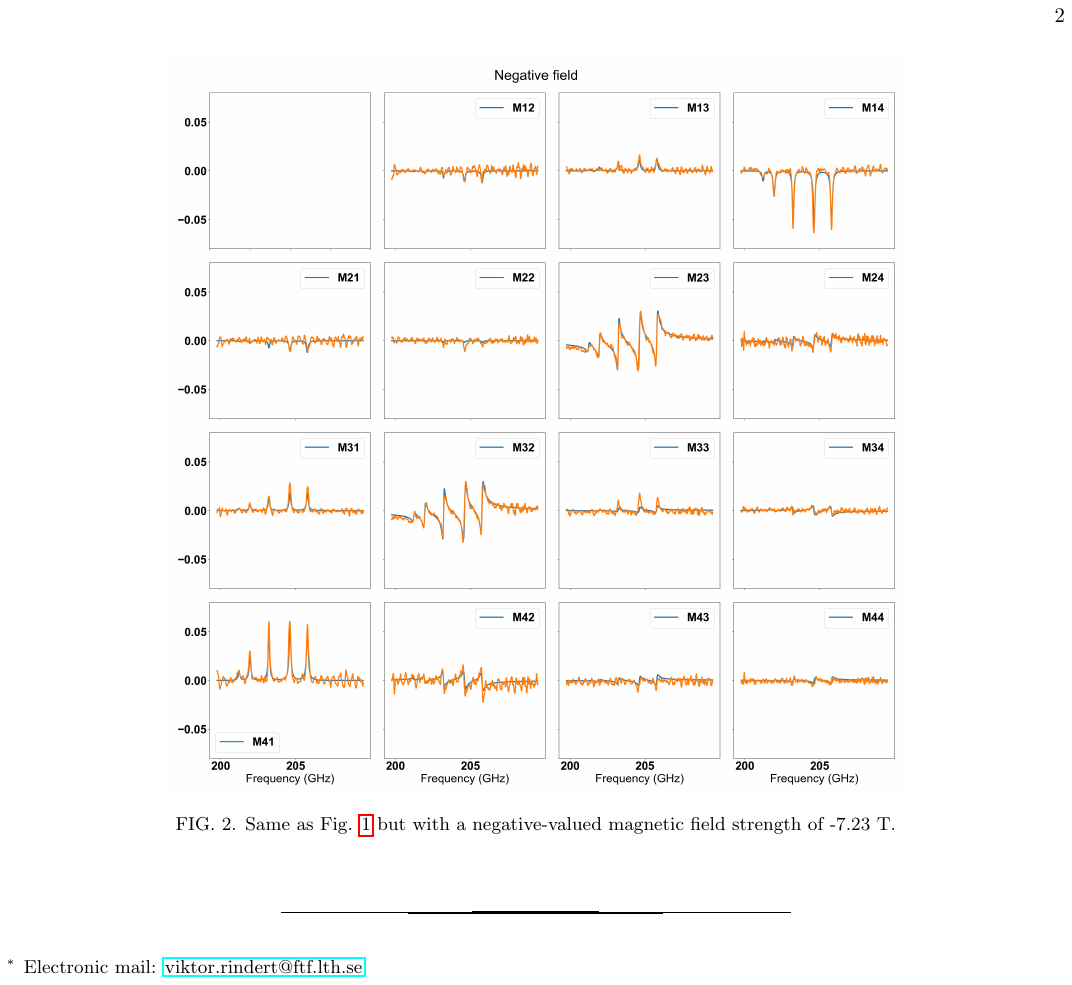}
\end{figure*}

\end{document}